\documentclass[11pt,preprint]{aastex}

\usepackage{amsmath}

\usepackage{natbib}
\usepackage{epsfig}
\usepackage{graphicx}

\shorttitle{Dynamical Evolution of Black Holes in Clusters}
\shortauthors{Morscher et al.}

\begin{document}

\title{The Dynamical Evolution of Stellar Black Holes in Globular Clusters}
  
\author{Meagan Morscher\altaffilmark{1,2}, Bharath Pattabiraman\altaffilmark{1,3}, Carl Rodriguez\altaffilmark{1,2}, Frederic A.\ Rasio\altaffilmark{1,2}, and Stefan Umbreit\altaffilmark{1,2}}

\affil{Center for Interdisciplinary Exploration and Research in Astrophysics (CIERA), Northwestern University, Evanston, IL, USA}
  \affil{Department of Physics and Astronomy, Northwestern University, Evanston, IL, USA}
 \affil{Department of Electrical Engineering and Computer Science, Northwestern University, Evanston, IL, USA}

\email{m.morscher@u.northwestern.edu}

\begin{abstract}
Our current understanding of the stellar initial mass function and massive star evolution suggests that young globular clusters may have formed hundreds to thousands of stellar-mass black holes, the remnants of stars with initial masses from $\sim 20 - 100\, M_\odot$. 
Birth kicks from supernova explosions may eject some 
black holes from their birth clusters, but most should be retained. 
Using a Monte Carlo method we investigate the long-term dynamical evolution of globular clusters containing large numbers of stellar black holes. We describe numerical results for 42 models, covering a range of realistic initial conditions, including up to $1.6\times10^6$ stars. In almost all models we find that significant numbers of black holes (up to $\sim10^3$) are retained all the way to the present. This is in contrast to previous theoretical expectations that most black holes should be ejected dynamically within a few Gyr. The main reason for this difference is that core collapse driven by black holes (through the Spitzer ``mass segregation instability'') is easily reverted through three-body processes, and involves only a small number of the most massive black holes, while lower-mass black holes remain well-mixed with ordinary stars far from the central cusp. Thus the rapid segregation of stellar black holes does not lead to a long-term physical separation of most black holes into a dynamically decoupled inner core, as often assumed previously. Combined with the recent detections of several black hole X-ray binary candidates in Galactic globular clusters, our results suggest that stellar black holes could still be present in large numbers in many globular clusters today, and that they may play a significant role in shaping the long-term dynamical evolution and the present-day dynamical structure of many clusters.

\end{abstract}



\section{Introduction} \label{Intro}

Massive star clusters ($M\gtrsim10^4\, M_\odot$) should form $\sim 100\, -\, 1000$ stellar-mass black holes (BHs) 
through normal stellar evolution, and, as long as BH birth kicks are sufficiently low, 
most should be retained initially in the cluster 
\citep{Belczynski2006,Willems2005,Wong2012}. 
With masses of $\sim 10\,M_\odot$, the BHs quickly segregate toward the dense central region 
of the cluster where they interact dynamically to form binaries with either a normal
star or another compact remnant as a companion. These binaries can evolve to produce X-ray binaries (XRBs) or 
merging compact object binaries potentially detectable by future ground-based gravitational wave 
(GW) observatories (LIGO, VIRGO; \citealt{HarryLIGO2010, VIRGO2014}). These systems could
be found either inside clusters or in the field after being dynamically ejected. 
It is well known that the formation rate 
per unit mass of XRBs is orders of magnitude larger in massive clusters than it is in the field
(e.g., \citealt{Pooley2003}), which suggests that stellar dynamics must play an essential
role in producing XRBs in present-day clusters. 

For several decades, observations, theoretical arguments, and simulations have all suggested that old 
globular clusters (GCs) should have very few (perhaps $\sim1$) BHs remaining at present.  
While many XRBs had been discovered in Galactic GCs \citep{Grindlay2001}, they had all been 
clearly identified as accreting neutron stars (NS) \cite[see][and references therein]{Kalogera2004}.
Furthermore, there were no good candidates for BHs in extragalactic GCs (for a review of GC X-ray sources as of 2006, see \citealt{VerbuntLewin2006}).
The absence of BHs from GCs was explained with simple theoretical arguments based 
on the prediction that all BHs should rapidly concentrate 
near the cluster center through dynamical friction from the low-mass background stars 
\citep{Kulkarni1993, Sigurdsson1993}. Eventually the BHs would succumb to the 
so-called Spitzer ``mass-segregation instability'' \citep{Spitzer1969, Kulkarni1993, Watters2000} 
and form a very dense 
subsystem within the cluster core that consists primarily of BHs and is dynamically 
decoupled from the other stars. The small-$N$ sub-cluster of BHs has a very short relaxation 
time, so it should promptly undergo its own core collapse, begin to form hard binaries through three-body 
interactions, and subsequently eject single and binary BHs. This sub-system 
should then completely evaporate within at most a few Gyr, leaving behind a GC essentially 
devoid of BHs well before reaching the $\sim 10$~Gyr ages typical of Galactic GCs. 
Several other theoretical studies later confirmed these predictions through numerical simulations 
(e.g., \citealt{PortegiesZwart2000}, \citealt{OLeary2006}, \citealt{Banerjee2010}). 

Over the last few years, however, our understanding of BHs in dense star 
clusters has taken a dramatic turn. The old story began to change when the first BH XRB 
candidate was identified inside an old GC in the Galaxy NGC~4472 \citep{MaccaroneNature2007}. 
Several more BH candidates have subsequently been discovered in extragalactic GCs 
\citep[e.g.,][]{Barnard2011, Maccarone2011, Shih2010}. Recently, \cite{Strader2012} discovered \emph{two} 
BHs inside of the Galactic GC M22. These stellar BH candidates are the first ever to be identified in a 
Milky Way GC, as well as the first to be discovered through radio observations. By assuming that these 
systems are BH--white-dwarf (WD) binaries, Strader et al.\ were able to use published theoretical models by 
\cite{Ivanova2010} to estimate the fraction of present-day BHs in GCs that are actively accreting from a WD 
companion. They estimate that the detection of two accreting BHs in M22 implies a total 
number of $\sim 5-100$ BHs.  The same group recently found another BH in a different 
galactic GC, M62, also through radio observations \citep{Chomiuk2013}. Several additional candidates
may soon be added to this list (Strader, private communication).

On the theoretical side, several recent studies have provided hints that old clusters might actually 
be able to retain significant numbers of BHs. \cite{Mackey2008} used $N$-body simulations of 
clusters with BHs to explain the trend of increasing spread in core radius with cluster age that is 
observed in the Magellanic Clouds. They found that a population of retained BHs could 
provide a heat source for some clusters, offering a possible explanation for the observed spread 
in the radii of Magellanic Cloud clusters. In some of their models, significant numbers of BHs 
(as many as $\sim100$) were retained for $\sim10$~Gyr.
\cite{Sippel2013} presented a scaled-down direct $N$-body model of M22. At an age of 12~Gyr, 
their model contains 16 BHs (about 1/3 of the initially-retained population), which is consistent
with the prediction of \cite{Strader2012}. Our own preliminary Monte Carlo study by \cite{Morscher2013} suggested 
that some clusters may retain as many as hundreds of BHs for 12~Gyr.  The long-term 
survival of such a large number of BHs is explained by the fact that the BHs do not become Spitzer 
unstable on the whole, but instead the majority of the BHs remain well mixed with the rest of the cluster 
throughout the entire 12~Gyr evolution.

A very different study by \cite{Breen2013} focused on the evolution of two-component clusters consisting of
a population of BHs co-existing within a background cluster of low-mass stars. They provide analytic calculations as 
well as direct $N$-body simulations which both suggest that the flow of energy between the sub-cluster of BHs and
the rest of the stars is ultimately determined by the cluster as a whole. From this it follows that the rate of energy 
production in the BH subsystem, as well as its evaporation rate, is also regulated by the whole cluster. 
This implies that BHs can be retained for much longer than previously thought (i.e., for $\sim 10\, t_{\rm rh,i}$, 
where $t_{\rm rh,i}$ is the initial half-mass relaxation timescale) because their dynamical evolution happens on 
the evolutionary timescale of the whole cluster, as opposed to that of the BH subsystem. 
This suggests that the long-standing assumption that BHs 
actually decouple from clusters, which is the basis for the argument that old clusters should be deplete 
of BHs, may no longer hold true.

While the theoretical arguments presented in \cite{Breen2013} are interesting and suggestive, these 
two-component models cannot be directly compared to real GCs, which have a  broad spectrum of 
stellar and BH masses, as well as larger total cluster masses. Several more-realistic studies have 
now predicted the survival of at least some BHs \citep[e.g.,][]{Heggie2014, Mackey2008, Morscher2013, Sippel2013}, but there is still no definitive answer as to \emph{how many} might actually be hiding 
in old GCs at present, nor whether models that do retain many BHs will look like observed Galactic GCs. 
The answers to these questions can help to constrain the initial populations of BHs and BH kicks, both of 
which are still highly uncertain \citep{Farr2011, Janka2013, Repetto2012}. 
For these reasons, the topic of stellar BHs in clusters is worthy of further theoretical study.

In this paper, we present a large grid of Monte Carlo simulations of realistic, large-$N$, 
Milky-Way-like GCs and address the question of retention of BHs in clusters and 
the dynamical evolution of clusters with BHs. We are most interested in understanding
whether clusters can retain significant numbers of BHs all the way to present and still have observable
properties similar to the GCs in our own Galaxy. 
Our focus, therefore, is on clusters that 
initially retain most of the BHs that form, under the assumption that BHs
receive small birth kicks (compared to NSs; See Section~\ref{MethodOverview}).
This work has been made possible by the recent parallelization of our code,
which has provided the speed up necessary to simulate star clusters with up to 
$\sim 10^6$ stars, large populations of BHs, and realistic stellar physics \citep{Pattabiraman2013}.
The rest of this paper is organized as follows. In Section~\ref{Method} we give an overview of
our computational method, including a new comparison with a direct $N$-body study focusing on BHs. We provide the
initial conditions that we have used for our calculations in Section~\ref{ICs}. The results of our 42 simulations
are described in detail in Section~\ref{Results}. Finally in Section~\ref{DiscussionConclusions} we summarize our results,
compare to previous studies, discuss the uncertainties in our assumptions, and give our conclusions.


\section{Monte Carlo Method} \label{Method}

\subsection{Overview of Method} \label{MethodOverview}

We use a Monte Carlo (MC) method for modeling the dynamical evolution of GCs.
While the direct $N$-body method is more accurate than MC schemes, 
it can only simulate clusters with up to $N \sim 10^5$ due to the
poor scaling with $N$ (computation time $\sim N^3/\log N$). In order to model large MW GCs
with initial $N$ up to $\sim 10^6$, and to cover the large parameter space of relevant initial conditions,
we must employ a more approximate technique. In MC methods, the computation time scales as 
   $\sim N \log N$, which makes it feasible to model realistic
 GCs and to study the evolution of rare objects, such as BHs.

Our MC implementation is a variation of the so-called ``orbit-averaged Monte Carlo 
method" developed by \cite{Henon1971a} for solving the Fokker-Planck 
equation. The details of our method are described in many previous studies
\citep{Joshi2000, Joshi2001, Fregeau2003, Fregeau2007, Chatterjee2010, Umbreit2012}
where we have also shown our results to be in excellent agreement with direct
$N$-body simulations.
Here we highlight the most important details for
our study of BHs in clusters. We treat the cluster on a star-by-star basis, which
makes it possible to layer on complexity, such as stellar evolution and strong binary 
interactions. Stars and binaries are evolved according to the stellar
evolution fitting formulae and interacting binary evolution calculations
of SSE and BSE (\citealt{Hurley2000}, \citealt{Hurley2002}). We use the 
modified stellar remnant formation prescription of \cite{Belczynski2002}, which
is based on the theoretical calculations of \cite{Fryer2001}. In this prescription,
BHs can form either through ``direct collapse" (i.e., with no supernova explosion)
or through partial fallback of material that was initially expelled in a supernova explosion,
depending on the mass of the stellar core just before BH formation. The range of initial masses
that form BHs, as well as the formation mechanism and the final BH masses, are all dependent
on the details of the stellar evolution scheme and metallicity.
For $Z=0.001$, our implementation produces BH masses in the range $\sim 3-30\, M_\odot$. 
Stars with initial masses $\gtrsim 25\, M_\odot$ directly collapse into BHs at the end of their lifetime,
while those with initial masses between $\sim 19-25\, M_\odot$ form BHs through fallback. 
All NSs and some BHs receive natal kicks
assumed to be generated by the asymmetric ejection of mass during a supernova explosion. 
NS kicks are drawn from a Maxwellian distribution with $\sigma=265$~km s$^{-1}$. 
We assume momentum-conserving kicks, which means that
BHs, being significantly more massive than NSs, receive much smaller kicks (if any).
We follow the prescription of \cite{Belczynski2002} to reduce the
BH kick magnitude (initially drawn from the NS kick distribution)
 according to the amount of material that falls back onto the 
final BH after the supernova explosion. In this prescription, 
BHs that form via direct collapse do not receive any natal kick, as there is no associated explosion.
For compact object binaries BSE calculates the orbital evolution due to 
emission of GW radiation, which is important for tracking the mergers of 
BH--BH binaries. Once a binary is ejected from the cluster, however, 
it is no longer evolved with our code, even though these BH--BH binaries can still 
potentially merge in the field. For these systems, we estimate the merger 
time using a simplified timescale for GW inspiral in the weak-field limit \citep{Peters1964} 
based on the
properties at the time of ejection.

In addition to two-body relaxation, it is also important to accurately model the dynamics 
of \emph{close} binary encounters. We choose strong binary-binary (B-B) and binary-single (B-S) 
using MC sampling based on the cross-section for a close interaction between the pair of 
neighboring objects. These interactions are then integrated directly using \texttt{Fewbody},
which allows for many important effects within binary systems, such as exchanges, ionization, 
hardening of binaries, and ejections, all of which are relevant for the evolution of BHs in clusters.


\subsection{New Physics: Three-body Binary Formation} \label{Method3bb}

We have recently implemented a simplified prescription for three-body 
binary formation, a process that is expected to produce an important 
population of hard BH binaries 
\citep{Kulkarni1993, Sigurdsson1993, PortegiesZwart2000,OLeary2006,  
Banerjee2010}, and is therefore extremely important for this study. 
If three single stars experience a close resonant encounter, it is possible for two
of the stars to become gravitationally bound to one another, with the third star 
carrying away the extra energy. The probability
of binary formation is usually quite low, and realistically only becomes significant 
under the extreme conditions expected at the core of a cluster which has been driven
to collapse by a population of BHs. For non-compact stars three-body 
binary formation is never important, as it would require a density so high that physical
collisions would instead have become dominant much earlier \citep{Chernoff1996}.
Therefore we restrict our attention to BHs. In addition, we are only interested in
dynamically hard binaries \citep{Fregeau2006, Heggie1975}, 
as only those are expected to survive within the cluster environment.

Our simplified prescription relies on the calculation of the rate at which 
three neighboring single BHs will form a hard binary. Using the calculated rate
and the current timestep, we can estimate the probability that the three-body
 system will result in binary formation, and then use MC sampling to select which systems
  will actually form a new binary. Our implementation follows
\citet{Ivanova2005}, \citet{Ivanova2010}, and \citet{OLeary2006}, where
the binary formation rate is expressed in terms of the binary hardness ratio
(binary binding energy to background star kinetic energy)
\begin{equation}
\eta = \frac{G \, m_1 \, m_2}{r_p \, \langle m \rangle \, \sigma^2}.
\label{eq:eta}
\end{equation}
Here $m_1$ and $m_2$ are the masses of the two stars assumed to
form a binary, $r_p$ is their separation at pericenter, and $\langle m \rangle$ and $\sigma$ are
the local average mass and velocity dispersion.

Keeping both the geometric and gravitational focusing contributions to
the cross section, we construct an expression for the
rate of binary formation for the selected neighboring three stars.
We calculate the rate at 
which two stars with masses $m_1$ and $m_2$ will form a binary with 
hardness $\eta \, \geq\, \eta_{\rm min}$ during a close encounter with a 
third star of mass $m_3$
using
\begin{multline}
\Gamma(\eta \geq \eta_{min}) = \sqrt{2} \pi^2 G^5 n^2
      {v_{\infty}^{-9}} \\ \times (m_1 + m_2)^5 \eta_{\rm min}^{-5.5} (1 + 2
      \eta_{\rm min}) \\ \times \left[ 1+2 \eta_{\rm min} \left( \frac{ m_1 + m_2 +
            m_3}{m_1 + m_2 } \right) \right],
\label{eq:Gamma}
\end{multline}
where $n$ is the local number density and $v_{\infty}$ is the average relative 
velocity at infinity, both of which are computed using a subset
of nearby stars (see Section~\ref{MethodNbody} for details).
We only form binaries with $\eta \geq 5 = \eta_{\rm min}$, with the specific
value chosen for each new binary from a distribution according to 
the differential rate, d$\Gamma$/d$\eta$, with lower limit $\eta_{\rm min}$. 
After a binary is formed, the new properties of all involved objects are calculated 
from conservation of momentum and energy.

\subsection{Comparison with Direct $N$-body Calculations} \label{MethodNbody}

The MC approach requires the calculation of local average of several
physical quantities.  For example, both the physics of three-body binary formation 
described above and the selection of the relaxation time depend on
the local number density, velocity dispersion, and average stellar mass
at a specific radius in the cluster \citep{Joshi2000}.
However, it is not the case that these averages should be computed over the same
number of stars.  While three-body binary formation should depend only on the
properties of neighboring stars, the relaxation time step must be applied to
 the entire cluster.
 We both expect and require three-body binary
formation to be more sensitive to local spikes in number density and velocity
dispersion than the cluster-wise relaxation time.  Therefore, we must adjust the
number of stars used for computing these averages, depending on the scale of the
physics in question.

\begin{figure}[h!]
  \epsscale{0.75}
	  \plotone{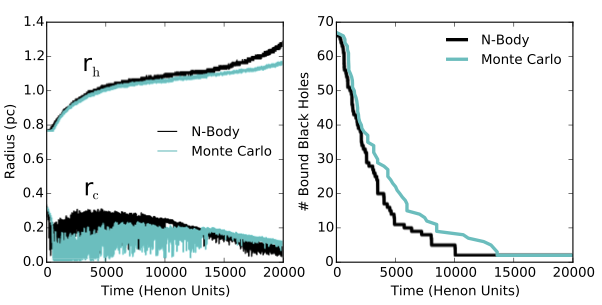}
 	 \caption{\scriptsize Evolution of two-component Plummer models as computed by our Monte Carlo
 	 code and the direct $N$-body code of \cite{Breen2013}.  On the left we 
	  show the half-mass radius, $r_{\rm h}$ (top), and core radius, $r_{\rm c}$ (bottom), for the
	  two methods, and on the right, the number of BHs retained in the cluster as
	  a function of time. With the choice of parameters  described in Section~\ref{MethodNbody}
	  we get good overall agreement with $N$-body results for the structural properties of our models, 
	  and also, crucially, on the BH ejection rate.}
  \label{fig:2species64k}
\end{figure}

As in previous works using our MC code 
\citep{Joshi2000, Joshi2001, Fregeau2003, Fregeau2007, Chatterjee2010}
we determine the optimal code
parameters by comparing to direct $N$-body simulations with identical initial
conditions.  Since the primary focus of this study is the retention of BHs,
we choose for our main test the idealized two-component models recently studied 
by \cite{Breen2013}, which provides a simplified description of the evolution of a population
of stellar-mass BHs in a cluster. These models are a realization of standard Plummer
spheres populated by a large population of low-mass stellar objects and a smaller
population of massive objects, considered to be BHs.  We consider models with an individual 
mass ratio of $m_2/m_1 = 20$, and a total cluster mass ratio of $M_2/M_1 = 0.02$, where 
$m_1$ and $m_2$ are the masses of individual particles, and $M_1$ and $M_2$ are the
total masses of each component.  We performed comparison simulations with $64k$ 
and $128k$ particles, although only the $64k$ runs are illustrated here.

In Figure~\ref{fig:2species64k}, we compare the cluster properties as reported by
the $N$-body simulations of \cite{Breen2013} to those computed by our MC code. 
Empirically, we find optimal agreement by computing the
average quantities over the nearest 40 stars for two-body relaxation, and the
nearest 6 stars for three-body binary formation.  In particular, the 
evaporation rate of the BH subcluster in our simulations matches
the $N$-body results very well.
Furthermore, we find that our MC
approach correctly reproduces the time evolution of the half-mass radius
to within 8\% after $2\times10^5$ $N$-body time units.  

Of the measured cluster properties, only the core radius cannot be
reproduced perfectly by the MC approach.  Immediately following core
collapse, the measured core radius for the MC model differs from the $N$-body
results by as much as 65\%. This is to be expected: once mass segregation and core collapse 
have occurred, the dynamics of the core is primarily driven by the BH subsystem, which has 
dynamically decoupled from the cluster in these idealized models. 
Modeling accurately the internal dynamics of a sub-system with $N\sim100$ particles
using an orbit-averaged MC approach is very challenging; however, as the BHs are
ejected, and the core becomes populated with a larger number of lower mass
stars, the validity of the MC approach is restored, and the core
radius better agrees with the direct $N$-body results.  Additionally, the core radius
is known to be very sensitive to stochastic physical effects, such as three-body binary formation,
so that the agreement between two different models can at best be statistical \citep{Giersz2008}.
New techniques are under investigation that will correctly evolve the subcluster dynamics while
maintaining the speed of the MC approach. For the present study,
we are encouraged by our earlier results presented in \cite{Morscher2013} 
which suggest that in \emph{realistic} GCs the BHs might actually never decouple 
from the cluster on the whole, in which case a MC approach is appropriate.


\section{Initial Conditions} \label{ICs}

Using the results of the calibration as described in Section~\ref{MethodNbody},
we have calculated the dynamical evolution of 42 cluster models with a wide range of
initial conditions. 
All models are initialized as King models \citep{King1966}, with 
stellar masses chosen in the range $0.1 \,-\, 100\, M_\odot$ according to the initial mass 
function (IMF) of \cite{Kroupa2001}, which is a broken power-law of the form 
$dN/dm \propto m^{-\alpha}$, with $\alpha=1.3$ for $0.08\, \le\,  m / M_\odot \,\textless \,  0.5$, 
and $\alpha=2.3$ for $m / M_\odot \,\ge \, 0.5$.
Once the single stars are drawn from the IMF, we randomly choose $N_{\rm b}$ stars to be the primary
member of a binary (where $N_{\rm b}$ is the total number of primordial binaries). The secondary masses
are drawn from a distribution that is uniform in the mass ratio within the range 
$0.1\, M_\odot \,-\, m_{\rm p}$, where $m_{\rm p}$ is the primary star mass.
The semi-major axes of the binaries are chosen from a distribution flat in $\log a$, where
the hardest binary has $a\, \textgreater\, 5\times (R_1 + R_2)$, where $R_1$ and $R_2$ are the
radii of the binary components, and the
softest binary is within the local hard-soft boundary (i.e., all primordial binaries are initially hard). 
The binary eccentricities are chosen from the thermal distribution (e.g., \citealt{HeggieHut2003}).

We vary the initial number of stars 
($N=2 \times 10^5$, $8 \times 10^5$, and $1.6 \times 10^6$), the initial King 
concentration parameter ($W_{\rm o}=2$, 5, 7) and the Galactocentric distance $R_{\rm G}$, 
which in our models corresponds to three different metallicities
($Z=0.005$ at $R_{\rm G}=2$ kpc, $Z = 0.001$ at $R_{\rm G}\,=\,8$ kpc and $Z=0.0001$ at $R_{\rm G}=20$ kpc). 
The choice to vary metallicity as a function of $R_{\rm G}$ was 
motivated by the observations of the Milky Way GC population, 
which show a correlation between $R_{\rm G}$ and $Z$, with larger metallicities 
being found closer to the Galactic center \citep{DjorgovskiMeylan1994}.
These initial conditions form a $3 \times 3 \times 3$ grid of 27 cluster models. 
Each of these models has initial virial radius $R_{\rm v}\,=\,2$ pc and binary fraction
$f_{\rm b} = 10$\%. We will call these 27 models our \emph{standard models}, and
name them according to the values of the three parameters $N$, $W_{\rm o}$, and $R_{\rm G}$ (e.g. \texttt{n8w5rg20} has $N=8 \times 10^5$,
$W_{\rm o}=5$, and $R_{\rm G}=20$ kpc, with metallicity $Z=0.0005$ set by $R_{\rm G}$).
 
 \begin{deluxetable}{lccccclcccc}
\tabletypesize{\tiny}
\tablewidth{0pc}
\tablecaption{Initial model parameters.}

\tablehead{
	\colhead{model} & \colhead{$N$} & \colhead{$M$} & \colhead{$W_0$} & \colhead{$R_{\rm v}$} & \colhead{$R_{\rm G}$} & \colhead{$Z$} & \colhead{$f_{\rm b}$} & \colhead{$r_{\rm c,dyn}$} & \colhead{$r_{\rm h,m}$} & \colhead{$\log_{10}(\rho_{\rm c})$}   \\
\colhead{} & \colhead{$(10^5)$} & \colhead{$(10^5 M_{\odot})$} & \colhead{} & \colhead{pc} & \colhead{(kpc)} & \colhead{} &  \colhead{\%} &\colhead{(pc)} &  \colhead{(pc)}  & \colhead{$(M_{\odot}$/pc$^3)$}
}

\startdata
n2w2rg2		& 2	& 1.36	& 2	& 2	& 2	& 0.005	& 10	& 1.0	& 1.7	& 4.47 \\ 
n2w2rg8		& 2	& 1.36	& 2	& 2	& 8	& 0.001	& 10	& 1.0	& 1.7	& 4.47 \\ 
n2w2rg20	& 2	& 1.36	& 2	& 2	& 20	& 0.0005	& 10	& 1.0	& 1.7	& 4.47 \\ 
n2w5rg2		& 2	& 1.36	& 5	& 2	& 2	& 0.005	& 10	& 0.7	& 1.6	& 4.75 \\ 
n2w5rg8$\,\dagger$		& 2	& 1.36	& 5	& 2	& 8	& 0.001	& 10	& 0.7	& 1.6	& 4.75 \\ 
n2w5rg20	& 2	& 1.36	& 5	& 2	& 20	& 0.0005	& 10	& 0.7	& 1.6	& 4.75 \\ 
n2w7rg2		& 2	& 1.36	& 7	& 2	& 2	& 0.005	& 10	& 0.4	& 1.6	& 5.25 \\ 
n2w7rg8		& 2	& 1.36	& 7	& 2	& 8	& 0.001	& 10	& 0.4	& 1.6	& 5.25 \\ 
n2w7rg20	& 2	& 1.36	& 7	& 2	& 20	& 0.0005	& 10	& 0.4	& 1.6	& 5.25 \\ 
n2-A		& 2	& 1.36	& 11	& 2	& 8	& 0.001	& 10	& 0.1	& 2.0	& 7.44 \\ 
n2-B$\,\dagger$		& 2	& 1.36	& 5	& 1	& 8	& 0.001	& 10	& 0.4	& 0.8	& 5.65 \\ 
n2-C		& 2	& 1.36	& 5	& 4	& 8	& 0.001	& 10	& 1.4	& 3.3	& 3.85 \\ 
n2-D		& 2	& 1.29	& 5	& 2	& 8	& 0.001	& 1	& 0.7	& 1.6	& 4.71 \\ 
n2-E		& 2	& 1.66	& 5	& 2	& 8	& 0.001	& 50	& 0.7	& 1.6	& 4.83 \\ 
\\ \hline \\
n8w2rg2		& 8	& 5.4	& 2	& 2	& 2	& 0.005	& 10	& 1.0	& 1.7	& 5.13 \\ 
n8w2rg8		& 8	& 5.4	& 2	& 2	& 8	& 0.001	& 10	& 1.0	& 1.7	& 5.13 \\ 
n8w2rg20	& 8	& 5.4	& 2	& 2	& 20	& 0.0005	& 10	& 1.0	& 1.7	& 5.13 \\ 
n8w5rg2		& 8	& 5.4	& 5	& 2	& 2	& 0.005	& 10	& 0.7	& 1.6	& 5.43 \\ 
n8w5rg8$\,\dagger$		& 8	& 5.4	& 5	& 2	& 8	& 0.001	& 10	& 0.7	& 1.6	& 5.43 \\ 
n8w5rg20	& 8	& 5.4	& 5	& 2	& 20	& 0.0005	& 10	& 0.7	& 1.6	& 5.43 \\ 
n8w7rg2		& 8	& 5.4	& 7	& 2	& 2	& 0.005	& 10	& 0.4	& 1.6	& 5.94 \\ 
n8w7rg8		& 8	& 5.4	& 7	& 2	& 8	& 0.001	& 10	& 0.4	& 1.6	& 5.94 \\ 
n8w7rg20	& 8	& 5.4	& 7	& 2	& 20	& 0.0005	& 10	& 0.4	& 1.6	& 5.94 \\ 
n8-A		& 8	& 5.4	& 11	& 2	& 8	& 0.001	& 10	& 0.1	& 2.0	& 8.08 \\ 
n8-B		& 8	& 5.4	& 5	& 1	& 8	& 0.001	& 10	& 0.4	& 0.8	& 6.34 \\ 
n8-C		& 8	& 5.4	& 5	& 4	& 8	& 0.001	& 10	& 1.4	& 3.3	& 4.53 \\ 
n8-D		& 8	& 5.13	& 5	& 2	& 8	& 0.001	& 1	& 0.7	& 1.6	& 5.39 \\ 
n8-E$\,\dagger$		& 8	& 6.57	& 5	& 2	& 8	& 0.001	& 50	& 0.7	& 1.6	& 5.51 \\ 
\\ \hline \\
n16w2rg2	& 16	& 10.82	& 2	& 2	& 2	& 0.005	& 10	& 1.0	& 1.7	& 5.38 \\ 
n16w2rg8	& 16	& 10.82	& 2	& 2	& 8	& 0.001	& 10	& 1.0	& 1.7	& 5.38 \\ 
n16w2rg20	& 16	& 10.82	& 2	& 2	& 20	& 0.0005	& 10	& 1.0	& 1.7	& 5.38 \\ 
n16w5rg2	& 16	& 10.82	& 5	& 2	& 2	& 0.005	& 10	& 0.7	& 1.6	& 5.67 \\ 
n16w5rg8	& 16	& 10.82	& 5	& 2	& 8	& 0.001	& 10	& 0.7	& 1.6	& 5.67 \\ 
n16w5rg20	& 16	& 10.82	& 5	& 2	& 20	& 0.0005	& 10	& 0.7	& 1.6	& 5.67 \\ 
n16w7rg2$\,\dagger$	& 16	& 10.82	& 7	& 2	& 2	& 0.005	& 10	& 0.5	& 1.6	& 6.18 \\ 
n16w7rg8	& 16	& 10.82	& 7	& 2	& 8	& 0.001	& 10	& 0.5	& 1.6	& 6.18 \\ 
n16w7rg20$\,\dagger$	& 16	& 10.82	& 7	& 2	& 20	& 0.0005	& 10	& 0.5	& 1.6	& 6.18 \\ 
n16-A	& 16	& 10.82	& 11	& 2	& 8	& 0.001	& 10	& 0.1	& 2.0	& 8.32 \\ 
n16-B	& 16	& 10.82	& 5	& 1	& 8	& 0.001	& 10	& 0.4	& 0.8	& 6.58 \\ 
n16-C	& 16	& 10.82	& 5	& 4	& 8	& 0.001	& 10	& 1.4	& 3.3	& 4.77 \\ 
n16-D	& 16	& 10.28	& 5	& 2	& 8	& 0.001	& 1	& 0.7	& 1.6	& 5.65 \\ 
n16-E	& 16	& 13.19	& 5	& 2	& 8	& 0.001	& 50	& 0.7	& 1.6	& 5.77 \\ 

\tablecomments{Initial conditions for all 42 models. Columns are as follows: model name, 
number of stars ($N$), total cluster mass ($M$) in $M_{\odot}$, King concentration parameter 
($W_{\rm o}$), virial radius ($R_{\rm v}$) in pc, galactocentric distance ($R_{\rm G}$) in kpc, 
metallicity ($Z$), binary fraction (\%), theoretical (mass-density weighted) core radius ($r_{\rm c,dyn}$) in pc 
\citep{Casertano1985}, theoretical half mass radius ($r_{\rm h,m}$) in pc, initial central 3D mass density in ($\log_{10}(\rho_{\rm c})$) 
in $M_{\odot}$/pc$^3$. 
The six models with a dagger by their
name indicate the representative  models that we have chosen to illustrate in several of the figures.
}
\enddata
\vspace{-0.5cm}
\label{table:initial_conditions}
\end{deluxetable}

We have also run fifteen additional models
in which we have either extended the range of one of the parameters
varied in the standard models, or varied a new parameter. For each $N$, starting with our 
intermediate parameters ($W_{\rm o}$ = 5, $R_{\rm G}$=8 kpc, $Z$=0.001),
we have created models with larger central concentration ($W_{\rm o}$ = 11), with smaller and
larger initial binary fraction ($f_{\rm b}$ = 1\% and 50\%), and with smaller and larger virial radius 
($R_{\rm v}$ = 1, 4 pc). At a given $N$, these models are designated with the letters A through E (e.g., \texttt{n2-A})
representing  $W_{\rm o}=11$ (A), $R_{\rm v}=1$ pc (B), $R_{\rm v}=4$ pc (C), $f_{\rm b}=1\%$ (D) and $f_{\rm b}=50\%$ (E).
Rather than attempting to reproduce the distribution of cluster properties
observed in the MW GCs, our goal is to see whether GCs with many BHs can 
evolve into $\sim 10$~Gyr old clusters that are consistent with the properties of MW GCs. 
We evolve all of our models to a final time\footnote{Three of our low-$N$ models evaporated
before 12~Gyr, ending the simulation early.} of 12~Gyr, which is a typical age for MW GCs. 
The properties of our initial models are given in Table~\ref{table:initial_conditions}.

For typical IMFs (e.g., Kroupa), a fraction of $\sim 10^{-4}\, - \,10^{-3} \, N$
stars should become BHs, depending on the exact mass range assumed for the IMF 
and the details of the stellar evolution assumptions
(e.g., the metallicity-dependent separation between NS and BH progenitors). 
For our low-, intermediate- and large-$N$ models, 
we form produce around 450, 1750, and 3500 BHs, respectively, which form from stars with initial 
masses above about $19\, M_\odot$. The BH mass spectrum depends significantly on the
metallicity assumed. In the \cite{Belczynski2002} remnant prescription used here the BH masses
range from $\sim 3-30\, M_\odot$ for $Z=0.0005$ and $Z=0.001$, but at 
higher metallicities ($Z=0.005$ here), mass loss from stronger stellar winds causes the upper 
end of the BH mass function to be truncated at about 20~$M_\odot$. For $Z=0.001$, about
36\% of the BHs are formed through partial fallback, and only these BHs receive natal kicks.
The rest of the BHs are formed through direct collapse.


\section{Results} \label{Results}

\subsection{Typical Evolution of Clusters with Black Holes} \label{ResultsOverview}

We start by describing the qualitative evolution common to all of our 
cluster models, and in later sections we describe in more detail the properties 
of the retained and ejected BH populations, as well as observable cluster properties. 
In what follows, whenever it is reasonable to show the results for all 
of our models we do so, but for cases when this is not possible, we have chosen 
three pairs of models (each with different $N$) that are identical except for one parameter, to
allow us to see the effect that the virial radius, the binary fraction, and the Galactocentric distance together 
with metallicity, have on our results. The pairs of models we selected are
 \texttt{n2w5rg8} and \texttt{n2-B} ($R_{\rm v}=2$~pc and $R_{\rm v}=1$~pc), 
\texttt{n8w5rg8} and \texttt{n8-E} ($f_{\rm b}=10\%$ and $f_{\rm b}=50\%$),
and \texttt{n16w7rg2} and \texttt{n16w7rg20} ($R_{\rm G}=2$~kpc with $Z=0.005$ and $R_{\rm G}=20$~kpc
with $Z=0.0005$).

\begin{figure}[!h]
\epsscale{0.9}
	\plotone{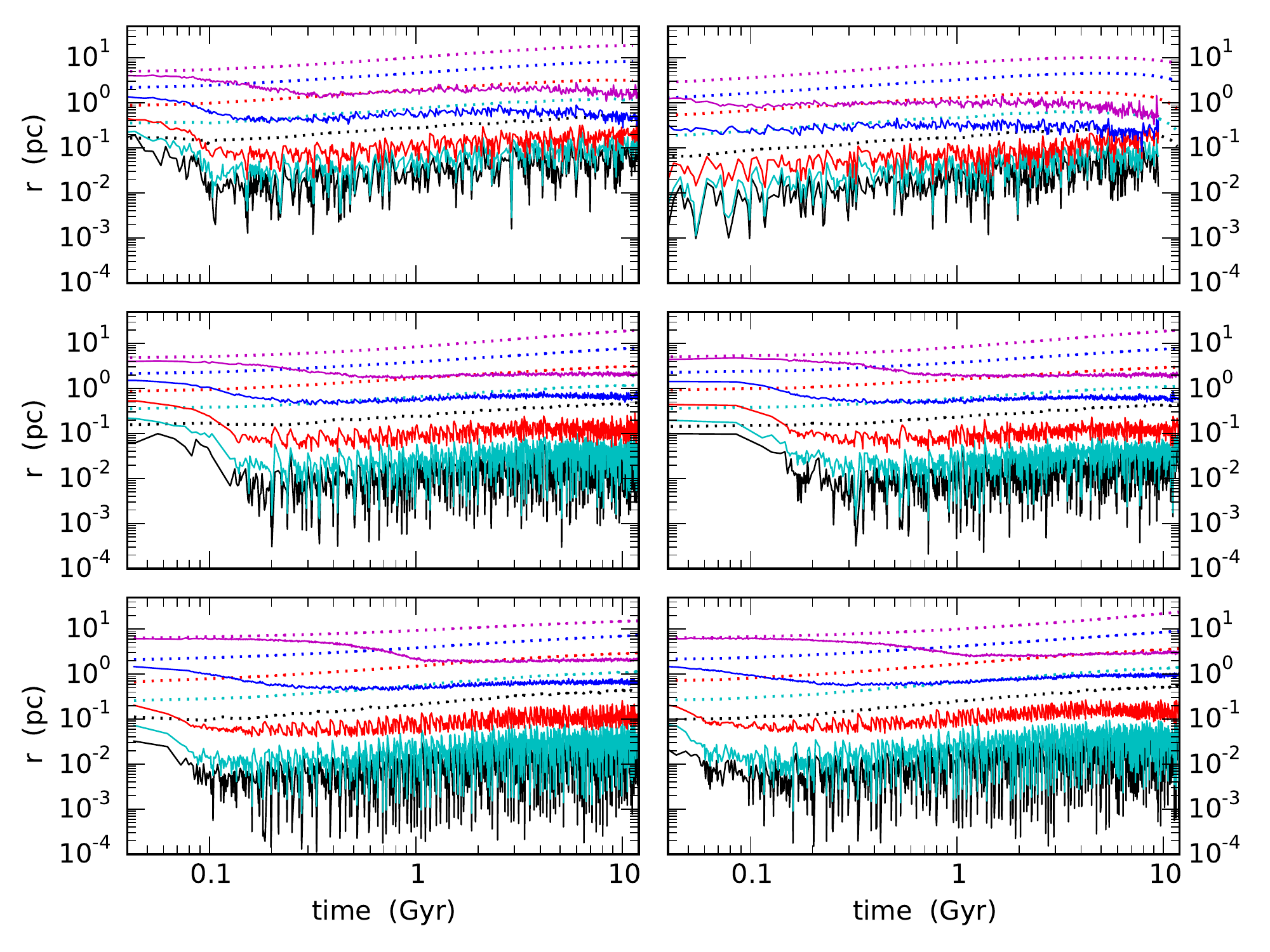}
	\caption{\scriptsize Evolution of the Lagrange radii for a subset of models, calculated separately 
	for the BHs (solid curves) and for all other objects (dotted curves). 
	The six models shown 
	are as follows: \emph{top left}:  \texttt{n2w5rg8}; 
	\emph{top right}: \texttt{n2-B}; 
	\emph{center left}: \texttt{n8w5rg8}; 
	\emph{center right}: \texttt{n8-E}; 
	\emph{lower left}:  \texttt{n16w7rg2}; 
	 \emph{lower right}: \texttt{n16w7rg20}. 
	The Lagrange
	radii shown enclose a fixed fraction of the mass (from bottom to top) of 0.1\%, 1\%, 
	10\%, 50\% and 90\%, for each individual component (BHs, non-BHs). 
	The central $\approx 1\%$ BH mass collapses within $\approx 100$ Myr, and the rest of 
	the BH mass segregates on a slightly longer timescale, while most of the rest of the cluster 
	steadily expands. 
	After a few Gyr, the $90\%$ BH Lagrange radius (solid magenta curve) typically crosses 
	inside of the $10\%$ radius for the rest of the cluster (dotted red curve). Model \texttt{n2-B}
	(top right) actually starts to contract near the end.
	}
	
	\label{fig:lagrad}
\end{figure}

Most of the BHs form within about 10~Myr and promptly begin to sink due to dynamical 
friction against the lower-mass background stars. 
The timescale for segregation of a BH from the half-mass radius to the core is 
\begin{equation}
t_{\rm seg} \sim \frac{\langle m \rangle}{m_{\rm BH}}\, t_{\rm rh} \sim 100\, \rm{Myr}
\label{eq:segregation timescale}
\end{equation}
\citep{OLeary2006}
where $\langle m \rangle$ is the average stellar mass, $m_{\rm BH}$ is the mass of the BH, 
and $t_{\rm rh} \sim1$~Gyr 
is the half-mass relaxation time. Since this timescale is dependent
on $m_{\rm BH}$, the most massive BHs tend to sink the fastest, driving a central 
collapse\footnote{This is different from what is usually referred to as ``core collapse," 
which occurs on a much longer timescale and will be discussed later. 
Terms such as ``core collapse" and ``post-collapse" are used inconsistently in the literature, and
can mean very different things to different authors, especially theorists vs observers 
(but the meaning can even vary between theorists) \citep[see][]{Chatterjee2013}.}.
This can be seen in Figure~\ref{fig:lagrad}, 
which shows the Lagrange radii for the six representative models. By looking at the Lagrange 
radii separately for the BHs (solid lines) and for the non-BHs (dotted lines), we see a clear 
separation of these two populations, with the BHs becoming more centrally concentrated 
than the lower-mass stars. A small subset (about 1\%) of the BH mass goes through radial 
oscillations where the 1\% Lagrange radius can vary by as much as two orders of magnitude.
 The other 99\% of the BH mass remains confined to roughly the same region for all time
as the rest of the cluster slowly expands.
The 90\% BH Lagrange radius is typically at about $1-2$~pc, and coincides
roughly with the 10\% Lagrange radius for the other (non-BH) stars.

\begin{figure}[!h]
\epsscale{0.6}
	\plotone{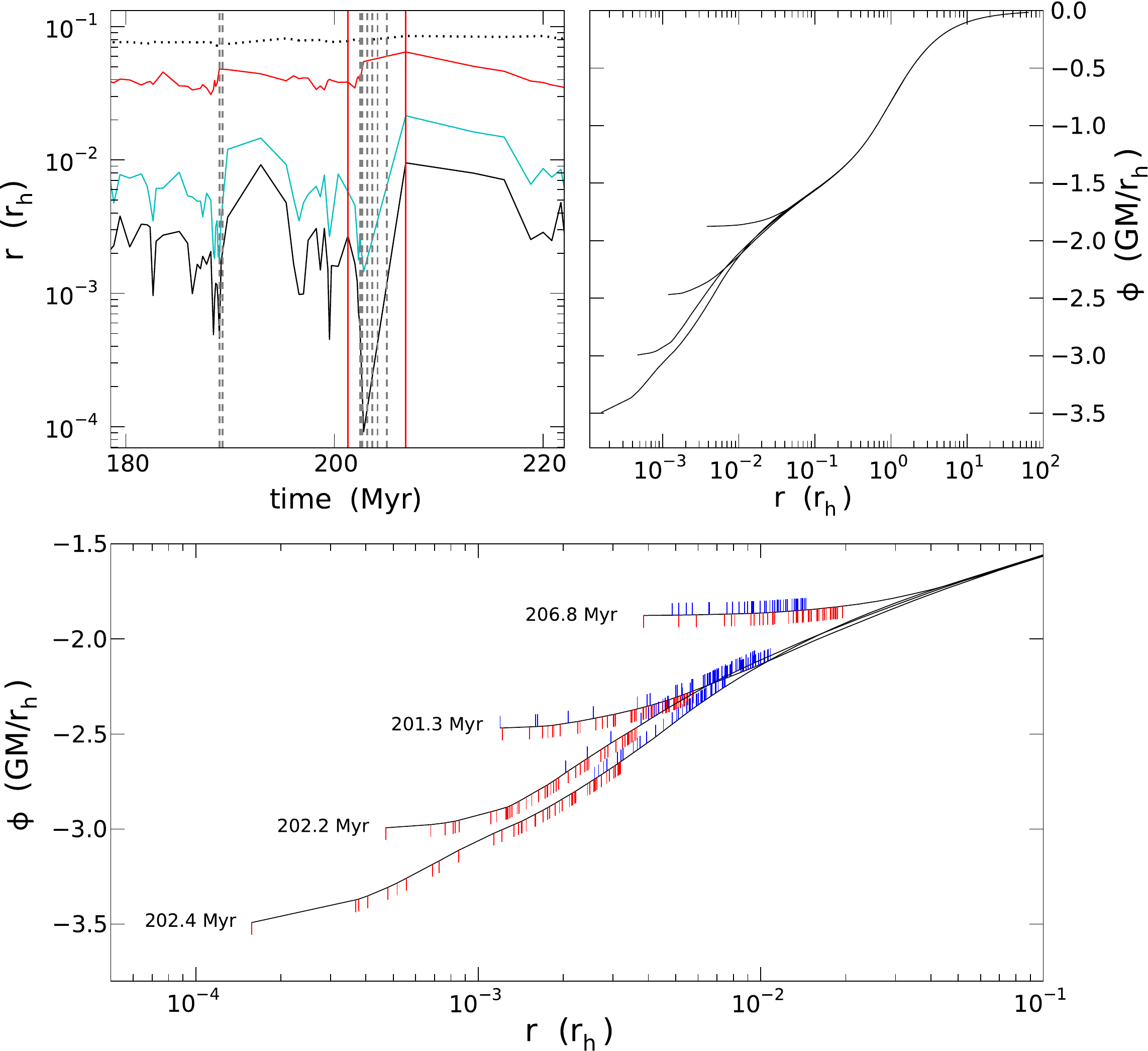}
	\caption{\scriptsize Time variation of central gravitational potential through a core oscillation
	around 200~Myr for model \texttt{n16w7rg20}
	(lower right panel in Figure~\ref{fig:lagrad}). 
	\emph{Top left}: zoom-in on the Lagrange radii from about 180--220~Myr.
	The radial coordinate is given in units of the initial half-mass radius 
	($r_{\rm h}$(0)). The three solid curves are the 0.1\%, 1\% and 10\% (\emph{from bottom to top}) Lagrange radii of the BHs, and the 
	dotted curve is the 0.1\% radius for all non-BH stars. The vertical dashed lines indicate the times
	when three-body binaries were formed. 
	The vertical solid red lines specify the period of time that we focus on in both the right 
	and the lower panels, which covers a deep collapse and subsequent re-expansion.
	\emph{Top right}: the full gravitational potential, $\phi(r)$, at four different times 
	(as indicated on the lower panel), in units of $G\, M / r_{\rm h}$, where $M$ is the
	total cluster mass and $r_{\rm h}$ is the half-mass radius, at that particular time. 
	\emph{Bottom}: zoom-in on the central potential, showing the radial positions of the innermost
	50 BHs (\emph{red ticks}) and non-BHs (\emph{blue ticks}) at each time.
	}
	\label{fig:phi}
\end{figure}

In Figure~\ref{fig:phi} we zoom-in on one of the core oscillations for model \texttt{n16w7rg20}
(lower right panel of Figure~\ref{fig:lagrad}) to show how 
the cluster potential fluctuates over a timescale of just a few Myr.
During the collapse of the innermost 1\% BH mass,
the BHs segregate from the lower-mass stars, forming a \emph{short-lived} cusp of mostly BHs. 
In the deepest part of the collapse, the central $\approx 30$ objects are \emph{all} BHs. Several 
three-body binaries form during this phase (see top left panel) and their interactions
with other objects ultimately power the re-expansion, after which the $1\%$ BH 
radius is even larger than it was pre-collapse.
At this point the BHs have become mixed with the other stars once again. 
These core oscillations occur frequently, anywhere from $\sim10-100$ times over 12~Gyr, 
depending on the model (see Figure~\ref{fig:lagrad}). The frequency and depth of the oscillations 
both depend on $N$ (i.e., deeper and more frequent for larger $N$). The oscillations also tend
decrease in frequency and become more shallow over time. Following the initial phase of BH segregation 
and rapid ejection (up to about a Gyr), the number of oscillations per Gyr decreases from as many as a 
few tens per Gyr (from 1--2 Gyr) down to just a few per Gyr near the end.

\begin{figure*}[!t]
\epsscale{0.7}
\tiny
	\plotone{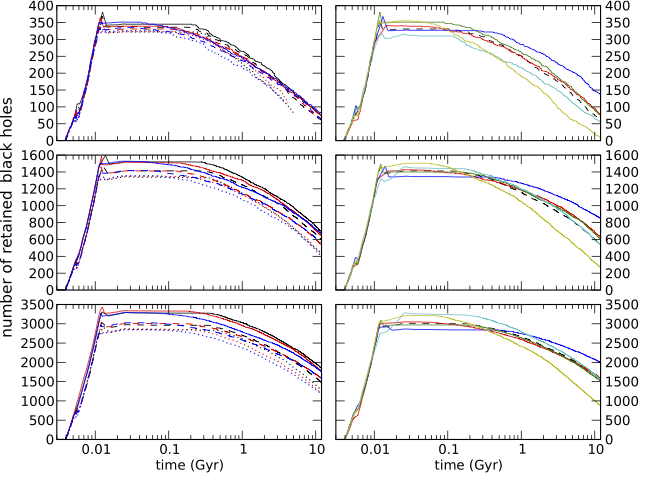}  
	\caption{\scriptsize Evolution of total number of retained BHs for all models.
	 From top to bottom, the adjacent panels show all models with initial $N=2 \times 10^5$, 
	 $N=8 \times 10^5$ , $N=1.6 \times 10^6$. On the left we show the nine 
	 standard models at each $N$. The color indicates the initial central 
	 concentration ($W_{\rm o}=2$ in black, $W_{\rm o}=5$ in red, and $W_{\rm o}=7$ in blue), the
	 linestyle indicates the initial galactocentric distance and metallicity (solid lines for
	 $R_{\rm G}=20$ kpc and $Z=0.0005$, dashed lines for $R_{\rm G}=8$ kpc and $Z=0.001$,
	 and dotted lines for $R_{\rm G}=2$ kpc and $Z=0.005$. All models on the left have 
	 $R_{\rm v}=2$ pc and $f_{\rm b}=10$\%. 
	 On the right we show the five additional models at each $N$, along with
	 the standard model for that $N$ with intermediate parameters 
	 (i.e., $W_{\rm o}=5$, $R_{\rm G}=20$ kpc, $Z=0.0005$, $R_{\rm v}=2$ pc, $f_{\rm b}=10$\%) for comparison
	 (black dashed curve).
	 Each of the solid colored lines has one parameter slightly different from this intermediate 
	 model, as follows: $W_{\rm o}=11$ shown in red, $R_{\rm v} = 1\,$pc and 4~pc in yellow and blue,
	 respectively, and $f_{\rm b}=1$\% and 50\% in green and cyan, respectively.
	 The three models on the top left that end before reaching 12~Gyr (dotted lines)
	 are clusters at $R_{\rm G}=2$ kpc that evaporated.
	 }

	\label{fig:ret_bhs_time}
\end{figure*}

We show the evolution of the total number of BHs present in each model in 
Figure~\ref{fig:ret_bhs_time}. We see most of the BHs forming up to about 10~Myr, as expected, and 
then after about 100~Myr, once the most massive BHs have segregated, the number of BHs
starts to decrease as they are ejected through strong binary encounters in the core.
The number of BHs continues to decrease all the way to 12~Gyr, but the rate slows down over time.
The majority of our low-, intermediate-, and large-$N$ models end with roughly 50--100 BHs,
400--800 BHs, and 1000--2000 BHs, respectively. While larger-$N$ models have more BHs at
12~Gyr than lower-$N$ models, they also eject a greater number of BHs in total.

\begin{figure*}[!h]
\epsscale{0.7}
	\plotone{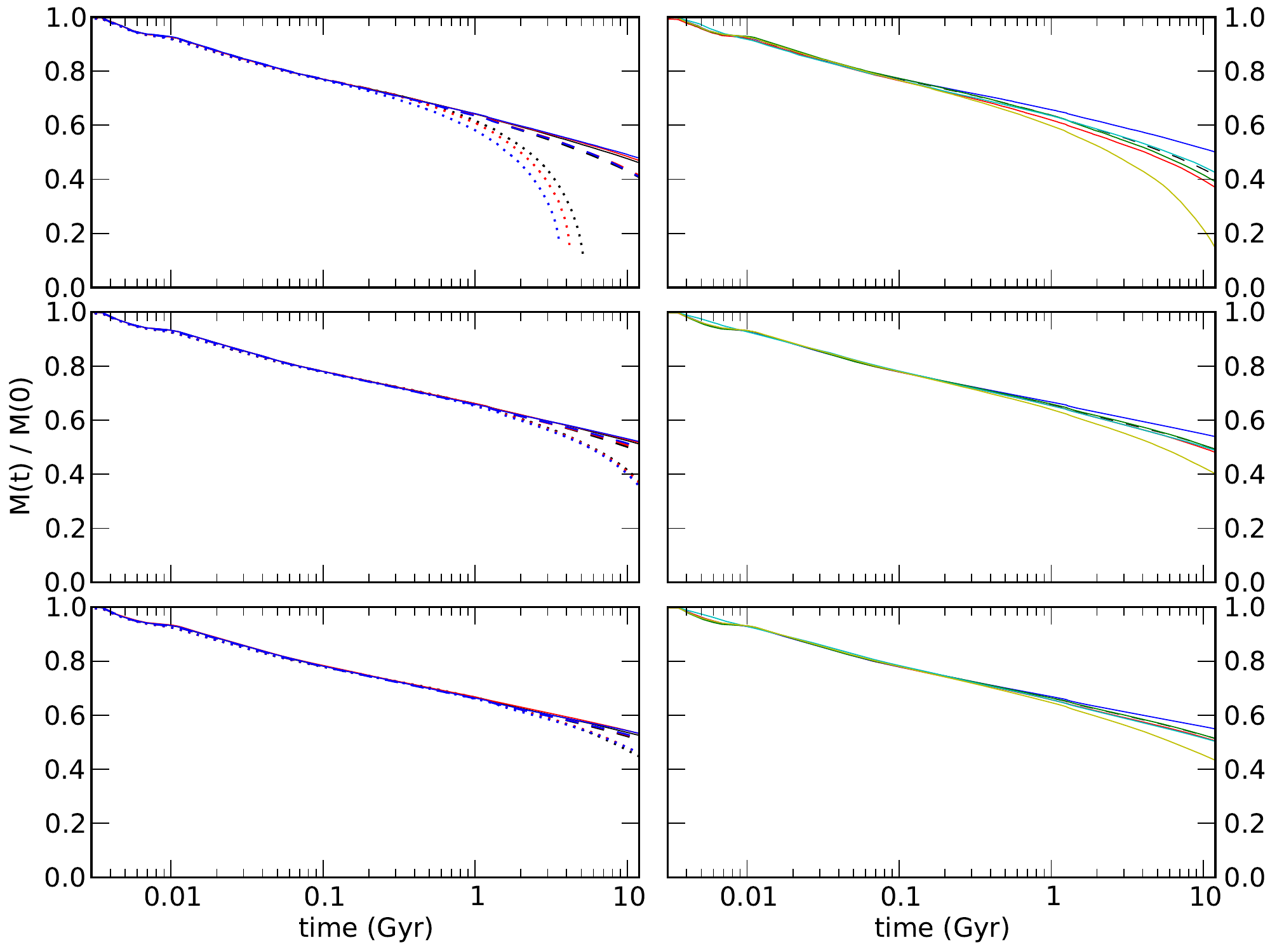}
	\caption{\scriptsize Evolution of the total cluster mass for all models. The various curves are described in Figure
	\ref{fig:ret_bhs_time}. Nearly all models lose about half of their mass over 12~Gyr.
	Models with smaller $R_{\rm G}$ (dotted lines on left panels) lose mass at a faster rate, and the smallest 
	of these models actually evaporate prior to 12~Gyr (top left panel). On the right panels, the virial
	radius has the greatest impact on mass loss, with the models having the smallest initial virial radii (yellow curves)
	losing mass at a faster rate than the rest, which is most evident for $N=2 \times 10^5$ (top right).
	}
	\label{fig:mass_time}
\end{figure*}

Figure~\ref{fig:mass_time} 
shows the evolution of total cluster mass for all models.  
After a period of rapid mass loss driven by early stellar evolution
of massive stars, the cluster mass loss rate tends to slow continually over time. 
Most of our models lose about half of their mass by 12~Gyr, but the most compact
clusters (with initial $R_{\rm v}=1\,$pc) and those at the smallest Galactocentric distances 
($R_{\rm G}=2$~kpc), which have the smallest tidal radii, lose mass at faster rates. 
In fact, among our low-$N$ models, the three with $R_{\rm G}=2$~kpc (\texttt{n2w5rg2, n2w5rg2, n2w7rg2})
nearly completely evaporate within about 6~Gyr (dotted lines in the upper left panel of Figure~\ref{fig:mass_time}),
and the model with $R_{\rm v}=1$~pc (\texttt{n2-B}) has lost more than 80\% of its mass by the end of the simulation. 
 The mass loss rate does not change significantly over the range $R_{\rm G}=8-20$~kpc.
 The final structural properties for all of our models are shown in Table~\ref{table:final_properties}.
 Note that these are all \emph{theoretical} properties (e.g., the density and core radius are 
 computing using all objects, not just luminous stars that can actually be observed). Observable
 properties of our clusters are discussed later.

 \begin{deluxetable}{lcrrrcrrr}
\tablecolumns{9}
\tabletypesize{\tiny}
\tablewidth{0pc}
\tablecaption{Final cluster properties. }

\tablehead{
\colhead{model} & \colhead{$N$} & \colhead{$M$} & \colhead{$r_{\rm c,dyn}$} & \colhead{$r_{\rm h,m}$} & \colhead{$\log_{10}(\rho_{c})$} & \colhead{$f_{b}$} & \colhead{$f_{\rm b,core}$}  \\
\colhead{ } & \colhead{$(10^5)$} & \colhead{$(10^5 M_{\odot})$} & \colhead{(pc)} & \colhead{(pc)} & \colhead{$(M_{\odot}$/pc$^3)$} & \colhead{\%} & \colhead{\%}
}

 \startdata
n2w2rg2	& 0.04	& 0.03	& 0.0	& 2.6	& 6.2	& 15.5	& 16.7 \\ 
n2w2rg8	& 1.43	& 0.55	& 2.9	& 7.8	& 2.64	& 9.5 	& 12.0 \\ 
n2w2rg20	& 1.65	& 0.63	& 3.2	& 8.6	& 2.54	& 9.3 	& 11.8 \\ 
n2w5rg2	& 0.02	& 0.03	& 0.4	& 1.6	& 3.84	& 13.0 	& 6.0 \\ 
n2w5rg8$\,\dagger$	& 1.46	& 0.56	& 3.1	& 8.3	& 2.95	& 9.5	& 12.3 \\ 
n2w5rg20	& 1.68	& 0.64	& 3.1	& 8.9	& 3.08	& 9.3 	& 11.3 \\ 
n2w7rg2	& 0.03	& 0.03	& 0.3	& 2.0	& 4.71	& 13.2	& 6.3 \\ 
n2w7rg8	& 1.44	& 0.56	& 3.5	& 8.9	& 2.74	& 9.5	& 12.3 \\ 
n2w7rg20	& 1.72	& 0.65	& 3.7	& 9.8	& 2.54	& 9.3 	& 12.0 \\ 
n2-A	& 1.28	& 0.5	& 3.8	& 9.4	& 2.69	& 9.5 	& 11.4 \\ 
n2-B$\,\dagger$	& 0.37	& 0.2	& 0.5	& 2.9	& 4.76	& 12.3	& 25.8 \\ 
n2-C	& 1.79	& 0.68	& 6.0	& 13.3	& 1.88	& 9.4 	& 10.9 \\ 
n2-D	& 1.36	& 0.51	& 3.6	& 8.6	& 2.37	& 1.0	& 1.0 \\ 
n2-E	& 1.54	& 0.71	& 2.7	& 7.7	& 3.31	& 46.5	& 53.7 \\ 
\\ \hline \\
n8w2rg2	& 5.12	& 2.04	& 2.5	& 6.2	& 3.91	& 9.3 	& 10.9 \\ 
n8w2rg8	& 6.86	& 2.62	& 2.2	& 7.9	& 5.23	& 9.1	& 10.2 \\ 
n8w2rg20	& 7.25	& 2.76	& 3.3	& 8.5	& 3.61	& 9.0	& 9.7 \\ 
n8w5rg2	& 4.99	& 2.0	& 2.7	& 6.6	& 3.69		& 9.4	& 10.7 \\ 
n8w5rg8$\,\dagger$	& 7.0	& 2.66	& 3.2	& 7.9	& 3.42	& 9.1 	& 10.1 \\ 
n8w5rg20	& 7.36	& 2.79	& 3.2	& 8.6	& 3.80	& 9.4 	& 10.0 \\ 
n8w7rg2	& 4.77	& 1.92	& 2.8	& 6.9	& 4.03	& 9.4 	& 11.1 \\ 
n8w7rg8	& 7.11	& 2.7	& 3.4	& 8.6	& 3.45	& 9.1		& 10.0 \\ 
n8w7rg20	& 7.41	& 2.81	& 0.0	& 9.3	& 8.98	& 9.0	& 4.8 \\ 
n8-A	& 6.77	& 2.6	& 2.8	& 9.0	& 4.74	& 9.1	& 9.7 \\ 
n8-B	& 5.56	& 2.17	& 1.7	& 4.9	& 4.28	& 9.1	& 11.9 \\ 
n8-C	& 7.6	& 2.91	& 3.9	& 11.7	& 4.44	& 9.2	& 9.7 \\ 
n8-D	& 6.87	& 2.52	& 3.0	& 8.3	& 4.29	& 1.0	& 1.0 \\ 
n8-E$\,\dagger$	& 7.19	& 3.21	& 2.9	& 7.7	& 3.93	& 45.1 	& 47.5 \\ 
\\ \hline \\
n16w2rg2	& 12.38	& 4.84	& 1.4	& 6.4	& 6.47	& 9.1	& 10.0 \\ 
n16w2rg8	& 14.39	& 5.5	& 1.6	& 7.5	& 6.16	& 8.9	& 9.3 \\ 
n16w2rg20	& 14.82	& 5.68	& 3.0	& 8.0	& 4.32	& 8.9	& 9.2 \\ 
n16w5rg2	& 12.77	& 4.97	& 2.1	& 6.6	& 5.34	& 9.1	& 10.1 \\ 
n16w5rg8	& 14.54	& 5.56	& 2.4	& 7.8	& 5.42	& 9.0	& 9.5 \\ 
n16w5rg20	& 14.82	& 5.76	& 0.0	& 8.5	& 10.1	& 8.9	& 25.0 \\ 
n16w7rg2$\,\dagger$	& 12.79	& 4.96	& 2.9	& 7.1	& 4.08	& 9.1	& 10.0 \\ 
n16w7rg8	& 14.61	& 5.58	& 2.8	& 8.4	& 4.98	& 9.0	& 9.3 \\ 
n16w7rg20$\,\dagger$	& 15.11	& 5.76	& 2.9	& 8.8	& 4.69	& 8.9 	& 9.0 \\ 
n16-A	& 14.23	& 5.47	& 3.3	& 8.5	& 3.93	& 8.9	& 9.2 \\ 
n16-B	& 12.17	& 4.69	& 1.5	& 5.1	& 5.85	& 9.0	& 10.5 \\ 
n16-C	& 15.46	& 5.94	& 4.8	& 11.1	& 3.67	& 9.1	& 9.4 \\ 
n16-D	& 14.38	& 5.28	& 2.9	& 7.8	& 4.48	& 1.0	& 0.9 \\ 
n16-E	& 14.8	& 6.63	& 3.0	& 7.6	& 4.04	& 44.7	& 45.0 \\

\tablecomments{Columns are as follows: model name, number of stars ($N$), total mass 
($M$) in $M_{\odot}$, theoretical (mass-density weighted) core radius ($r_{\rm c,dyn}$) in pc 
(which is very different from that which an observer would measure; 
see Section~\ref{ResultsObservables} and Table~\ref{table:Observables}), 
half-mass radius ($r_{\rm h,m}$) in pc, central 3D mass density in ($\log_{10}(\rho_{\rm c})$) in $M_{\odot}$/pc$^3$, final overall binary fraction ($f_{\rm b}$), and final binary fraction in the core ($f_{\rm b,core}$), as defined above. 
All properties are calculated at $t=12$ Gyr, except for models 
\texttt{n2w2rg2, n2w5rg2 and n2w7rg2}, which evaporated prior to 12~Gyr. 
For these models, the properties are calculated at the time when we deemed the cluster to 
have almost completely evaporated (when there are only about 1000 stars remaining), 
which happens at 5.2, 4.2, and 3.6 Gyr respectively for the models listed above.
As in Table~\ref{table:initial_conditions}, 
the six representative  models shown in several of the figures are marked with a dagger.
A core size of 0.0 (\texttt{n2w2rg2}, \texttt{n8w7rg20}, \texttt{n16w5rg20}) means that the BH core is in 
a collapsed state at the end of the simulation, so the core is extremely small and ill-defined. 
This is of course unrelated to the core radius that an observer would calculate.
The exceptionally small and large final core binary fractions in models \texttt{n8w7rg20} and \texttt{n16w5rg20} 
(respectively) has to do with their being in a collapsed state, where the core is composed
of a very small number of stars, and thus the binary fraction is quite sensitive to small fluctuations 
in the core composition.
}
\enddata
\vspace{-0.5cm}
\label{table:final_properties}
\end{deluxetable}


\begin{figure*}[!h]
\epsscale{0.7}
	\plotone{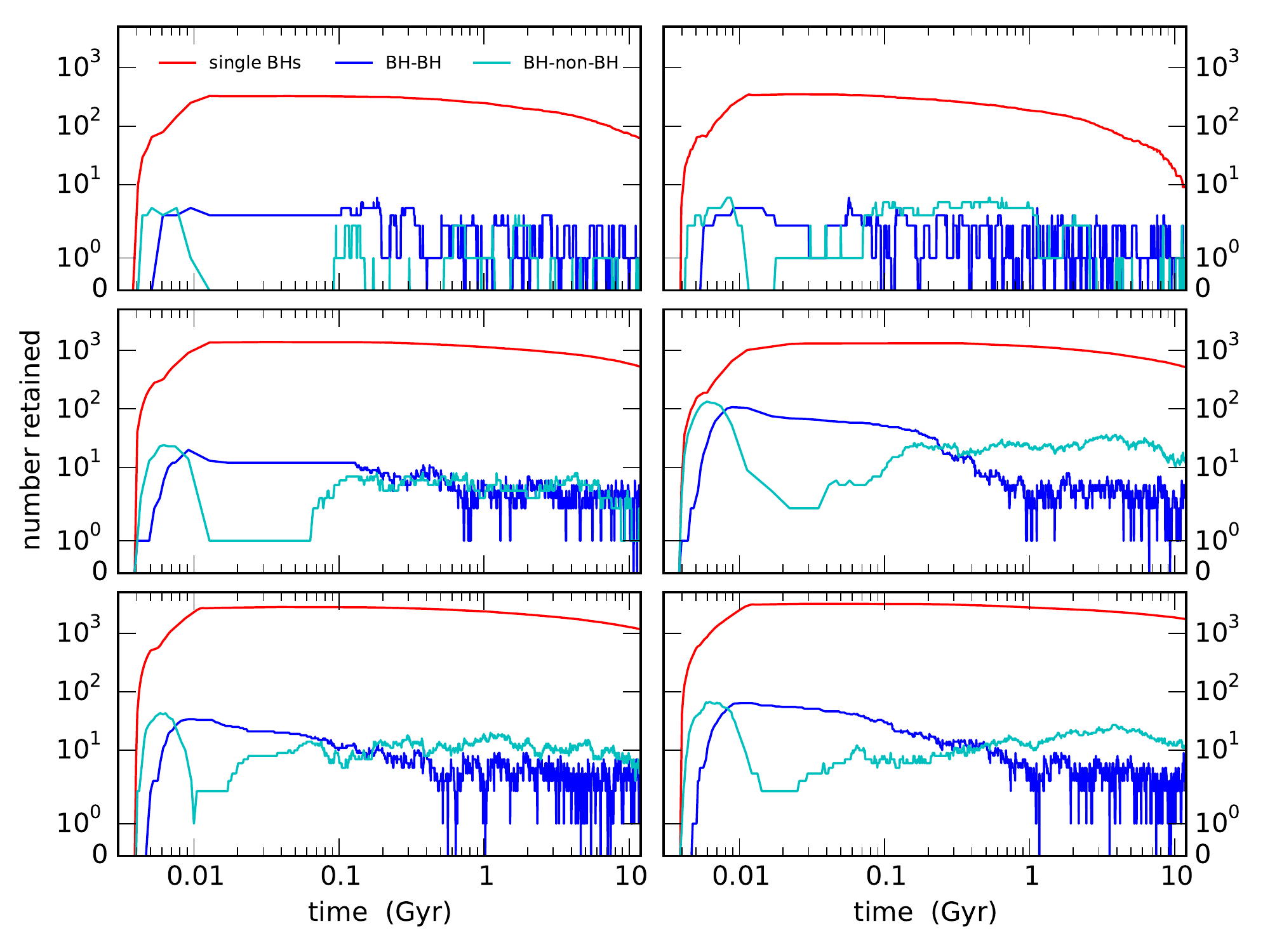}
	\caption{\scriptsize Numbers of \emph{retained} single and binary BHs as a function of time for 
	the six models shown in Figure~\ref{fig:lagrad}. Numbers of single BHs are in red, BH--BH binaries
	in blue, and BH--non-BH binaries in cyan. 
	Nearly all retained BHs are single. 
	With increasing $N$ (top to bottom) the number of BHs in binaries increases slightly.
	A larger binary fraction (center right, $f_{\rm b}=50\%$, compared to 10\% on left) allows more BHs
	to be in binaries (mostly BH--non-BH), but does not have a significant impact on overall BH retention.
	The number of BH binaries is not affected significantly by either virial radius 
	(compare top panels) or $R_{\rm G}$  and $Z$ (compare lower panels).
	}
	
	\label{fig:ret_bhs_binaries}
\end{figure*}

\subsection{Retained Black Hole Populations} \label{ResultsRetained}

\begin{figure*}[!ht]
\epsscale{0.7}

	\plotone{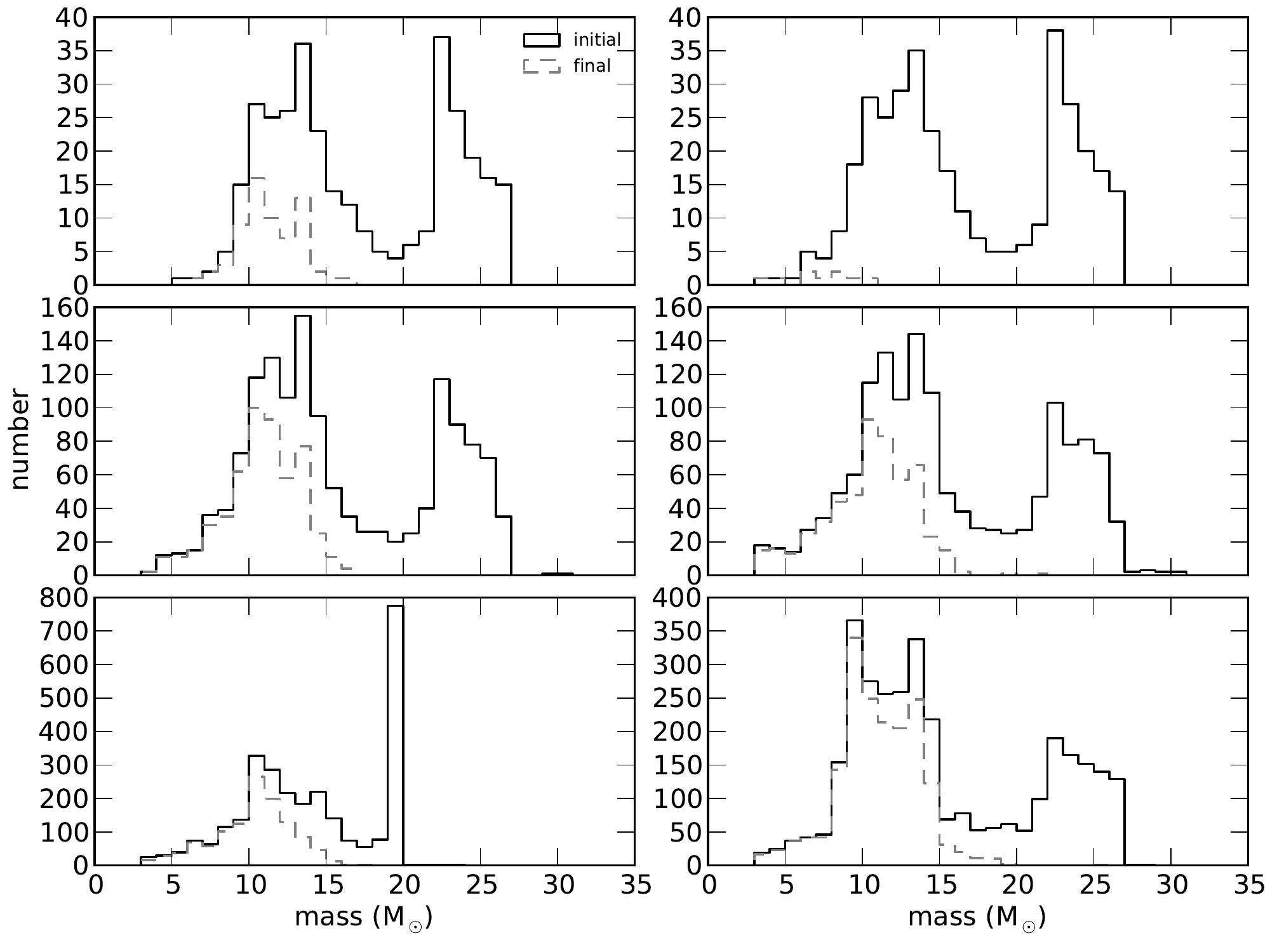}
	\caption{\scriptsize  BH mass spectrum initially (i.e., at a time between $30-100\,$Myr) (solid black lines) 
	and at 12~Gyr (dashed grey lines) for the same six models described in Figure~\ref{fig:lagrad}.
 	The `initial' BH mass spectrum is based on the BHs in the cluster at an early time 
	(around 30-100 Myr, depending on model).
 	The most massive BHs are always the first to be ejected, which reduces the
	initially double-peaked BH mass spectrum to a single peak. 
	The maximum BH mass and location of the peak depends on the
	fraction of BHs that have been ejected. 
	The initial BH mass spectrum looks very different for high $Z$
	(compare lower panels, with $R_{\rm G}=2$ kpc and $Z=0.005$ on the left and 
	$R_{\rm G}=20$ kpc and $Z=0.0005$ on the right; note that the different y-axis scales 
	are different). For large $Z$, mass lost from stellar wind (most significant for massive stars)
	prevents the most massive BHs (above $20\,M_\odot$) from forming, and causes
	the pileup of BHs at about $20\,M_\odot$, which results from a flattening
	of the progenitor-to-remnant mass relation (see Figure 1 of \citealt{Belczynski2004}).
	Model \texttt{n2-B} (top right) has $R_{\rm v}=1\,$pc and ejects nearly all of its BHs, 
	leaving behind just 9 BHs with masses of $3-10\, M_\odot$.
	}

	\label{fig:bh_mass_functions}
\end{figure*}

Next we look at the properties and evolution of the retained BHs in more detail and
discuss differences among our models.
The initial BH mass spectrum is shown in Figure~\ref{fig:bh_mass_functions}, and aside from 
the normalization, the only factor that significantly
affects the mass function is the metallicity $Z$. Since massive and metal-rich stars lose 
more mass via stellar winds, they form less massive BHs than do lower metallicity stars 
(see lower right panel).  
Our models retain between $65-90\%$ of the BHs \emph{initially}, depending primarily
on $R_{\rm G}$ (and $Z$) and $R_{\rm v}$. The reason for the $R_{\rm G}$ and $Z$ dependence 
of the initial retention fraction is twofold: First, a BH with a given position and kick speed 
will escape more easily from the cluster with the smaller tidal radius. Additionally, since 
models with smaller $R_{\rm G}$ also have larger $Z$, the BHs produced have lower
masses and will therefore tend to receive stronger kicks, making these objects even 
more likely to be ejected upon formation. More compact clusters (small $R_{\rm v}$) 
can retain initially formed BHs more easily.

In Figure~\ref{fig:ret_bhs_binaries} we show the distribution of single and binary 
BHs as a function of time for our six representative models. Here we see that almost all of the retained BHs
remain as single stars throughout the cluster evolution, in agreement with our earlier results \citep{Morscher2013}.
There are usually no more than a few tens of BH binaries of any type
inside the clusters at any given time, and are usually made up of comparable 
numbers of BH--BH and BH--non-BH binaries. 
A larger supply of primordial binaries does provide more opportunities for BHs to exchange into binaries 
through dynamical interactions and so we see a slightly larger number of BH binaries in models with larger $f_{\rm b}$. 
This effect can be seen in the center panels in Figure~\ref{fig:ret_bhs_binaries}, where we 
compare model \texttt{n8-E} ($f_{\rm b}= 50\%$, right) to model \texttt{n8w5rg8} ($f_{\rm b}=10\%$, left).
Since most of the primordial binary population
consists of two low-mass stars initially (which will never become BHs), the number of  
BH--non-BH binaries is most affected by the primordial binary fraction. 
The other parameters seem to have only a minor effect on the number of BH binaries in clusters.

The final retained BH mass distributions are shown in Figure~\ref{fig:bh_mass_functions}
along with the initially retained population, for comparison. 
Since the most massive BHs segregate the deepest they also interact the most frequently, and 
therefore tend to be the first to be ejected. Over time,
the maximum BH mass in the cluster is reduced from about $25-30\, M_\odot$ initially down to 
about $15-20\, M_\odot$ at 12~Gyr. Many of our models still contain a
substantial population of $\approx 10\, M_\odot$ BHs at 12~Gyr. 
The fraction of (initially retained) BHs that are retained all the way to 12~Gyr depends strongly on $N$.
For our largest-$N$ clusters, the final retention fraction $f_{\rm{BH},12}$ is typically about 50\%, and for the
lowest-$N$ clusters $f_{\rm{BH},12}$ is only about 20\% (except for the special case of model \texttt{n2-B}, 
which we will discuss separately). Since the initial number of BHs \emph{and} the final BH retention fraction
both scale with $N$, the final \emph{number} of BHs grows faster than linearly with $N$.
The final properties of the populations of retained and ejected BHs for each model are given in 
Table~\ref{table:bhs}.

Looking back to Figure~\ref{fig:ret_bhs_time} we see that
models with smaller $R_{\rm G}$ retain fewer BHs at 12~Gyr, but this is primarily because
they retained fewer BHs \emph{initially}.
Although larger primordial binary fractions produce a slightly larger 
number of BHs in binaries, this has little impact on the final number of retained BHs. For example,
comparing models with $N=1.6 \times 10^6$ and $f_{\rm b}=1\%$ and 50\% (\texttt{n16-D} and \texttt{n16-E}),
the final number of retained BHs at 12 Gyr is 1512 and 1556, respectively (see Table~\ref{table:bhs} for details). 
Rather, it seems that three-body binaries play a much more significant role in overall BH evaporation 
(although, as we will show later, the binary fraction does impact the number of ejected 
BH--non-BH binaries, as well as the number of in-cluster BH--BH mergers).
By far, the initial virial radius has the greatest impact on the BH evaporation rate for
models with a given $N$.
More compact clusters with smaller $R_{\rm v}$ are more dense, 
and so they process their BHs at a faster rate and therefore end with significantly fewer
BHs (compare the yellow and blue curves on each panel on the right hand side of 
Figure~\ref{fig:ret_bhs_time}). Furthermore, since the massive BHs are depleted to a greater 
extent, the remaining population is composed of BHs with comparatively low masses.
The model with the fewest BHs remaining at 12~Gyr is 
\texttt{n2-B}, our low-$N$ model with $R_{\rm v}=1$ pc,
which has just 9 BHs at 12~Gyr, or $f_{\rm{BH},12}$= 2\% (compared to 135 BHs for model
\texttt{n2-C} with $R_{\rm v}=4$ pc, but same initial conditions otherwise). 
We see a similar trend in BH retention in our more massive cluster models, 
but the contrast is not quite as stark (269 BHs retained in model \texttt{n8-B} 
compared to 852 in model \texttt{n8-C}; 869 BHs retained in model \texttt{n16-B}
 compared to 1988 in model \texttt{n16-C}).
It seems that the only way to get rid of most or all of the BHs is to start with very small $R_{\rm v}$.


\begin{deluxetable}{l@{\hskip 0.2 in}c@{\hskip 0.1 in}|@{\hskip 0.1 in}r@{\hskip 0.15 in}r@{\hskip 0.1 in}c@{\hskip 0.0 in}c@{\hskip 0.05 in}r@{\hskip 0.18 in}|@{\hskip 0.1 in}r@{\hskip 0.1 in}rrc@{\hskip 0.0 in}r@{\hskip 0.2 in}|@{\hskip 0.1 in}r}

\tabletypesize{\tiny}
\tablewidth{0pt}
\renewcommand{\tabcolsep}{0.1cm}
\tablecaption{Numbers of BHs retained in and ejected from each cluster.}
\tablehead{
\colhead{} & \colhead{initial} & \colhead{total} & \colhead{single} & \colhead{BH--BH} & \colhead{BH-WD} & \colhead{BH-star\, } & \colhead{total} & \colhead{single} & \colhead{BH--BH} & \colhead{BH-WD} & \colhead{BH-star}  & \colhead{mergers} \\

\colhead{model} & \colhead{formed/ret} &  \multicolumn{5}{l}{\rule{1.3cm}{.1pt}  \, final retained \,  \rule{1.3cm}{0.1pt}} & \multicolumn{5}{l}{\rule{1.3cm}{.1pt}  \, final ejected \,  \rule{1.3cm}{0.1pt}} & \colhead{ret / ej}\\
}

\startdata
n2w2rg2 &430 / 322      & 81    & 76    & 2     & 0     & 1     & 312   & 245   & 33    & 0     & 1     & 0 / 9 \\
n2w2rg8 &459 / 336      & 58    & 57    & 0     & 0     & 1     & 381   & 304   & 36    & 0     & 5     & 0 / 5 \\
n2w2rg20        &474 / 345      & 76    & 74    & 1     & 0     & 0     & 384   & 291   & 45    & 0     & 3     & 1 / 4 \\
n2w5rg2 &427 / 323      & 111   & 110   & 0     & 0     & 1     & 285   & 232   & 25    & 0     & 2     & 0 / 6 \\
n2w5rg8$\,\dagger$ &457 / 331      & 65    & 61    & 2     & 0     & 0     & 361   & 287   & 36    & 0     & 2     & 0 / 6 \\
n2w5rg20        &471 / 339      & 73    & 71    & 1     & 0     & 0     & 393   & 298   & 44    & 0     & 7     & 0 / 12 \\
n2w7rg2 &433 / 326      & 116   & 111   & 2     & 0     & 1     & 297   & 237   & 29    & 0     & 2     & 0 / 5 \\
n2w7rg8 &459 / 330      & 66    & 62    & 2     & 0     & 0     & 369   & 302   & 31    & 0     & 5     & 0 / 6 \\
n2w7rg20        &477 / 347      & 82    & 77    & 2     & 0     & 1     & 378   & 302   & 35    & 0     & 6     & 2 / 5 \\
n2-A    &464 / 338      & 76    & 72    & 1     & 1     & 1     & 368   & 290   & 37    & 0     & 4     & 0 / 4 \\
n2-B$\,\dagger$    &463 / 350      & 9     & 9     & 0     & 0     & 0     & 428   & 317   & 54    & 0     & 3     & 1 / 12 \\
n2-C    &454 / 327      & 135   & 128   & 3     & 0     & 1     & 309   & 251   & 25    & 0     & 7     & 1 / 1 \\
n2-D    &456 / 351      & 74    & 70    & 2     & 0     & 0     & 362   & 290   & 36    & 0     & 0     & 0 / 1 \\
n2-E    &472 / 310      & 55    & 51    & 1     & 0     & 2     & 386   & 286   & 41    & 0     & 18    & 8 / 18 \\
\\ \hline \\
n8w2rg2 &1689 / 1338    & 399   & 391   & 2     & 0     & 4     & 1181  & 923   & 127   & 0     & 3     & 8 / 58 \\
n8w2rg8 &1788 / 1413    & 598   & 586   & 3     & 1     & 5     & 1120  & 917   & 101   & 0     & 1     & 14 / 52 \\
n8w2rg20        &1813 / 1519    & 690   & 680   & 4     & 0     & 2     & 1056  & 822   & 116   & 0     & 2     & 14 / 57 \\
n8w5rg2 &1692 / 1349    & 429   & 422   & 1     & 0     & 5     & 1163  & 920   & 120   & 0     & 2     & 8 / 67 \\
n8w5rg8$\,\dagger$ &1779 / 1412    & 533   & 525   & 3     & 0     & 2     & 1182  & 926   & 127   & 0     & 2     & 12 / 65 \\
n8w5rg20        &1809 / 1512    & 643   & 634   & 4     & 0     & 1     & 1112  & 860   & 122   & 0     & 6     & 15 / 57 \\
n8w7rg2 &1698 / 1355    & 437   & 426   & 3     & 0     & 5     & 1154  & 891   & 130   & 1     & 2     & 5 / 69 \\
n8w7rg8 &1780 / 1417    & 562   & 553   & 4     & 0     & 1     & 1149  & 907   & 120   & 0     & 1     & 14 / 68 \\
n8w7rg20        &1815 / 1520    & 666   & 659   & 1     & 0     & 5     & 1109  & 837   & 133   & 0     & 6     & 17 / 68 \\
n8-A    &1790 / 1419    & 638   & 623   & 4     & 2     & 5     & 1098  & 874   & 105   & 1     & 10    & 0 / 57 \\
n8-B    &1809 / 1503    & 269   & 263   & 1     & 1     & 3     & 1461  & 1112  & 174   & 0     & 1     & 9 / 115 \\
n8-C    &1747 / 1346    & 852   & 837   & 4     & 2     & 5     & 857   & 708   & 68    & 0     & 10    & 11 / 25 \\
n8-D    &1749 / 1401    & 602   & 594   & 4     & 0     & 0     & 1104  & 901   & 101   & 0     & 1     & 0 / 51 \\
n8-E$\,\dagger$    &1949 / 1446    & 534   & 514   & 3     & 0     & 14    & 1262  & 922   & 157   & 0     & 24    & 56 / 97 \\
\\ \hline \\
n16w2rg2        &3477 / 2850    & 1261  & 1250  & 2     & 0     & 7     & 2050  & 1608  & 220   & 0     & 2     & 25 / 181 \\
n16w2rg8        &3634 / 2966    & 1473  & 1464  & 2     & 0     & 5     & 2048  & 1618  & 213   & 0     & 4     & 30 / 159 \\
n16w2rg20       &3737 / 3282    & 1848  & 1831  & 4     & 1     & 8     & 1801  & 1358  & 219   & 0     & 5     & 20 / 179 \\
n16w5rg2        &3458 / 2864    & 1202  & 1194  & 0     & 0     & 8     & 2099  & 1613  & 242   & 0     & 2     & 20 / 194 \\
n16w5rg8        &3659 / 3008    & 1585  & 1566  & 5     & 1     & 8     & 1974  & 1563  & 204   & 0     & 3     & 27 / 152 \\
n16w5rg20       &3841 / 3333    & 1770  & 1748  & 5     & 1     & 11    & 1951  & 1453  & 244   & 0     & 8     & 21 / 194 \\
n16w7rg2$\,\dagger$        &3447 / 2844    & 1176  & 1163  & 4     & 0     & 5     & 2078  & 1628  & 222   & 0     & 6     & 17 / 180 \\
n16w7rg8        &3638 / 3026    & 1587  & 1570  & 4     & 2     & 7     & 1949  & 1545  & 198   & 0     & 7     & 15 / 159 \\
n16w7rg20$\,\dagger$       &3721 / 3283    & 1757  & 1738  & 4     & 0     & 11    & 1867  & 1407  & 225   & 0     & 8     & 31 / 168 \\
n16-A   &3666 / 3043    & 1582  & 1559  & 7     & 0     & 9     & 2012  & 1606  & 197   & 0     & 7     & 0 / 156 \\
n16-B   &3703 / 3213    & 869   & 859   & 2     & 1     & 5     & 2710  & 2027  & 337   & 0     & 4     & 21 / 287 \\
n16-C   &3610 / 2852    & 1988  & 1967  & 4     & 1     & 12    & 1536  & 1242  & 145   & 0     & 3     & 30 / 98 \\
n16-D   &3548 / 2972    & 1512  & 1502  & 5     & 0     & 0     & 1954  & 1576  & 189   & 0     & 0     & 3 / 142 \\
n16-E   &4259 / 3250    & 1556  & 1497  & 5     & 4     & 45    & 2330  & 1771  & 269   & 1     & 17    & 156 / 235 \\

\enddata

\vspace{-0.5cm}
\tablecomments{The first two columns are the model name and the total number of BHs that formed/retained initially in each model; columns 3--7 give the total number of BHs, single BHs, BH--BH, BH--WD, and BH--star binaries \emph{retained} through the end of each simulation; similarly columns 8--12 give the total number of BHs, single BHs, BH--BH, BH--WD, and BH--star binaries \emph{ejected} by the end of each simulation. The final column shows the number of mergers that occur within the cluster/post ejection from the cluster.}
\label{table:bhs}

\end{deluxetable}


\subsection{Ejected Black Hole Binaries} \label{ResultsEjected}

\begin{figure*}[t]
\epsscale{0.7}
	\plotone{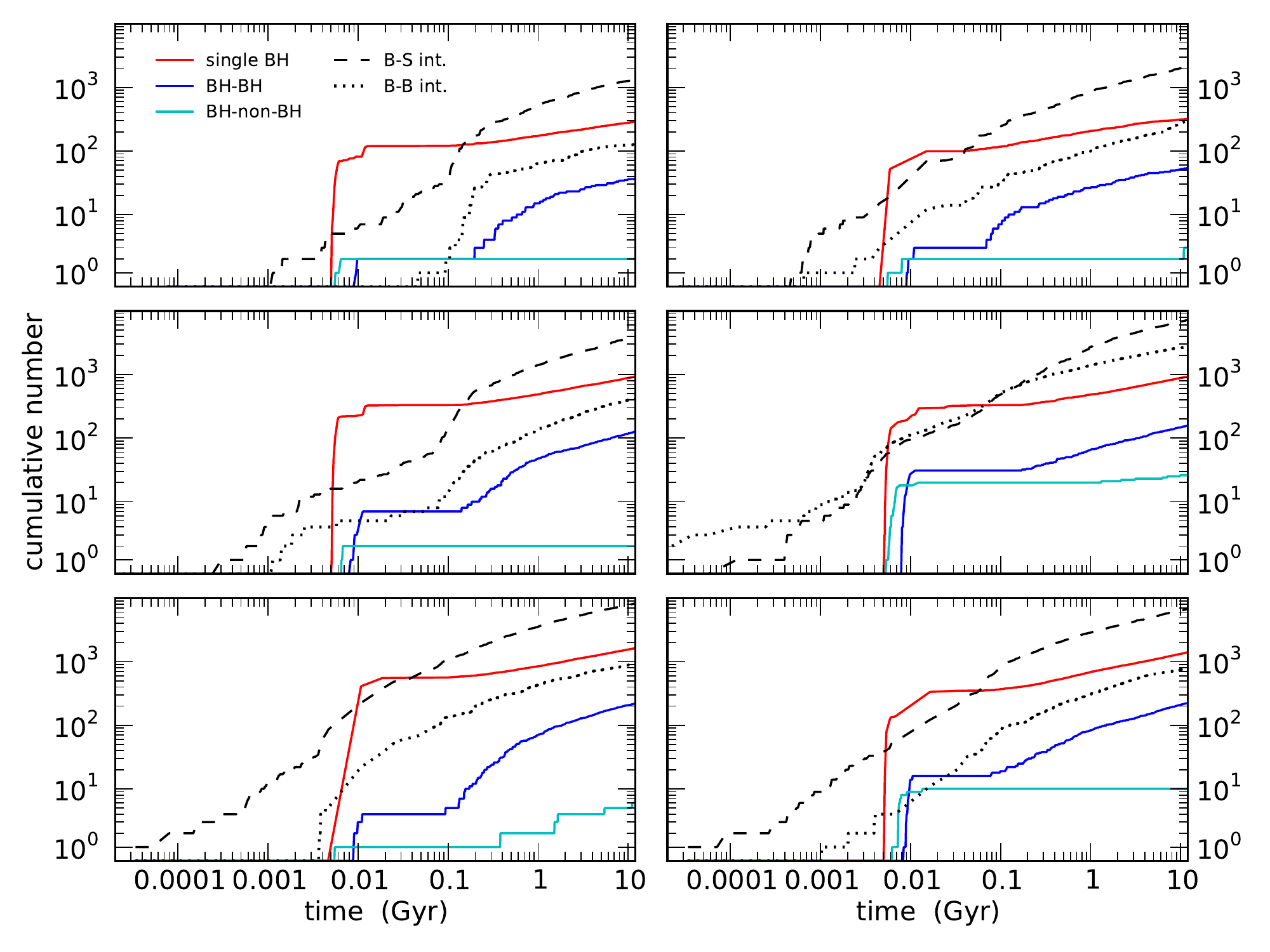}
	\caption{\scriptsize Cumulative number of binary interactions and ejected BHs as a function of time for the 
	six models shown in Figure~\ref{fig:lagrad}. Black dashed and dotted curves show 
	the cumulative number of binary-binary (B-B) and binary-single (B-S) interactions, respectively. 
	The solid curves show the cumulative number of ejected single BHs (red),
	BH--BH binaries (blue), and BH--non-BH binaries (cyan). 
	The sharp increase at around 10 Myr shows BHs that are ejected via natal kicks. 
	In many models, we see an increase in the binary interaction rate
	after about 100 Myr, and an associated increase in the BH--BH ejection rate.
	Most of the ejected BH binaries are BH--BH binaries. 
	Models with binary fraction $f_{\rm b}=50\%$ (center right) have similar B-S and B-B interaction
	rates,
	while all other models with lower binary fractions have many more single stars than binaries, 
	and hence have mostly B-S interactions.
	}
	
	\label{fig:binint}
\end{figure*}

We now examine the ejected BH populations with a focus on BH binaries.
In Figure~\ref{fig:binint} we show the cumulative number of binary-binary
(B-B) and binary-single (B-S) interactions along with the total number of ejected single BHs, 
BH--BH binaries, and BH--non-BH binaries as a function of time. 
The early phase of rapid 
BH ejection is associated with BHs that are ejected at birth via supernova kicks prior to 10 Myr, 
and is followed by a flattening of the BH ejection rate, during which time the BHs
are segregating. Once the BHs have segregated sufficiently, the dynamical BH ejections
begin, typically by about $100-300$ Myr. 
In more massive clusters with higher central densities, binary interactions 
begin much more gradually, but also earlier, so they have already become important
well before the BHs have formed. In the lower-$N$ models, binary interactions 
begin later, and in some cases the segregation of BHs actually drives an increase 
in the rate of binary interactions (e.g., top left and center left panels 
in Figure~\ref{fig:binint}).
As in the case of the retained BHs, the ejected BHs too are mostly single.
In order to eject a BH binary, it typically has to participate in multiple binary
interactions in order to harden enough that the recoil from some final interaction
is sufficient to remove it from the cluster entirely.
The binaries that are ejected, therefore, are most often BH--BH 
binaries, since through many strong interactions, any low-mass non-BH binary 
companions will be preferentially replaced by BHs. This holds true even
though there are usually about as many (and sometimes more) BH--non-BH binaries 
present in our models as there are BH--BH binaries.

We find that most of the (small number of ) BH--non-BH binary ejections happen 
early, at the time of formation of the BH, rather than through subsequent dynamics.
Most models have many more B-S interactions than B-B interactions, since with $f_{\rm b}=10\%$ 
there are many more single stars than binaries, but with $f_{\rm b}=50\%$, there are about equal
numbers of B-S and B-B interactions. In models with large binary fractions 
(e.g., center right panel in Figure~\ref{fig:binint}), the number of BH--non-BH 
ejections is greater than in models with fewer binaries, but they still occur primarily
at BH formation. There are slightly more dynamical BH--non-BH ejections with higher
binary fractions, since there is a greater supply of binaries with which
the BHs could interact (the same reason that there are also more BH--non-BH binaries 
present inside the clusters).
In the lower panels in Figure~\ref{fig:binint} we see that in 
the high metallicity cluster (left), fewer BH--non-BH binaries are ejected upon BH 
formation, and instead are mostly ejected dynamically. This makes sense considering
that the the lower-mass BHs produced at high metallicities will receive larger birth kicks,
which will unbind (rather than eject) more of these binaries when the BH is formed. Additionally, 
low-mass BHs will also be more likely to interact with normal low-mass stars over time, which explains 
why this model ejects more BH--non-BH binaries through dynamics than in some other models.

\begin{figure*}[!t]
\epsscale{0.9}

	\plotone{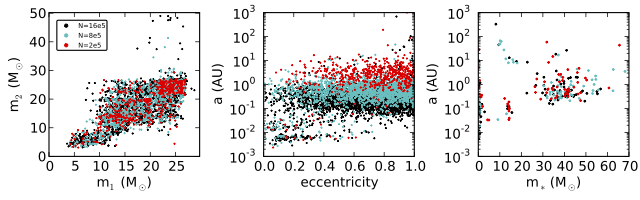}	
	\caption{\scriptsize Properties of ejected BH binaries at time of ejection from all simulations combined. Each point represents
	one BH binary, and the color indicates the initial $N$ for the cluster from which the binary originated
	(black for $N=1.6 \times 10^6$, cyan for $N=8 \times 10^5$, and red for for $N=2 \times 10^5$. 
	The first two panels show properties of the $\approx 5600$ ejected BH--BH binaries:
	on the left, $m_2$ versus $m_1$, and in the middle the semi-major axis versus the eccentricity. 
	The masses of the BHs in the ejected BH--BH binaries are similar across all $N$,
	while their orbital properties depend strongly on $N$, with more massive clusters forming and ejecting much tighter BH--BH binaries. 
	On the right we show the semi-major axis and the companion mass for the 227 ejected binaries 
	containing a BH with a non-compact stellar companion. 
	Note that the points are plotted on top of each other (from bottom to top: black, 
	then cyan, then red), so some points are hidden, especially in the left and center panels. The trends are
	visible, nonetheless.
	}
	\label{fig:bh_binaries}
\end{figure*}

The virial radius has the greatest effect of the ejection rate of BH--BH binaries. The
smaller $R_{\rm v}$, the higher the density and therefore also the binary interaction rate. We find that
for a given $N$, models with $R_{\rm v}=1$ pc eject more than twice the number of BH--BH binaries 
as models with $R_{\rm v}=4$.
 This trend does not hold for ejected BH--non-BH binaries. 
In fact, for $N=2 \times 10^5$ and $N=8 \times 10^5$, the models with smaller virial 
radii actually eject \emph{fewer} BH--non-BH binaries than those with larger virial radii. 
Since most of the BH--non-BH binaries are ejected at the time of BH formation rather than 
through dynamics, we should not expect the rate of these ejections to increase with the higher 
interaction rates occurring in more compact clusters.

In Figure~\ref{fig:bh_binaries} we show the binary properties for the ejected BH binaries
at time of formation, color-coded by the initial $N$ of the model from which it originated. 
On the first panel we show the masses of the components ($m_1$ and $m_2$) of
BH--BH binaries ejected from all models.
Since all clusters with a given metallicity form the same spectrum of BH masses, 
and the more massive BHs are ejected before the less massive ones, it is not surprising 
that the masses of the BHs in ejected BH--BH binaries is nearly independent of $N$.
In the \emph{orbital} properties of ejected 
BH--BH binaries (center panel), however, we see a very obvious correlation with $N$.
The least massive clusters eject binaries with significantly larger 
semi-major axes (typically $\sim 1$~AU) than the most massive clusters (typically 
$\sim 0.1$~AU). This follows from the fact that it is easier to eject a binary from a 
less massive cluster due to its lower escape speed, 
therefore most of these binaries get ejected before they have a chance to tighten
to sub-AU separations. On the right panel we show the semi-major axis and the companion 
mass for the 227 ejected binaries containing a BH with a non-compact stellar companion.
Recall that most of these binaries are  ejected within about 10 Myr, which is why the majority
of these systems have massive companions that have not yet evolved into compact
objects (note that once ejected, these objects are no longer evolved). There is no obvious trend 
with $N$, but there is a weak correlation between $a$ and $m_*$, 
with the binaries containing more massive companions tending to have slightly larger separations.

\subsection{Merging BHs} \label{ResultsMergers}

The dynamics that leads to BH evaporation also produces very tight BH--BH binaries, many of which
merge within the 12~Gyr.
These mergers can either occur while the binary is still bound to the cluster or in the field after 
being ejected. For all of our models combined, we produce 4096 merging systems over the 12~Gyr 
of evolution. Of these mergers, about 85\% occur post-ejection in the field and the other 15\% occur inside 
clusters. Nearly 71\% of the mergers are produced in the large-$N$ models, about 26\% in the 
intermediate-$N$ models, and just under 3\% in the lowest-$N$ models. The numbers of mergers 
per model are given in Table~\ref{table:bhs}. The strong $N$-dependence of the merger rate is 
caused by two effects. First, we saw that more massive clusters processed more BHs, and therefore 
ejected a greater number of BH--BH binaries. Also, since more massive clusters tend to eject 
\emph{tighter} binaries, it turns out that a greater fraction of the ejected binaries actually merge 
within 12~Gyr. The binary fraction has the next biggest impact of the merger rate, enhancing the 
rate of post-ejection mergers slightly, but dramatically increasing the rate of in-cluster mergers 
(going from $1\%$ to $50\%$ binaries, the number of in-cluster mergers for our models increases 
from 0 to 8, 0 to 56, and 3 to 156, in order of lowest to highest $N$). 
This means that it is not only dynamically formed hard three-body binaries that produce BH--BH mergers, 
but also BH--BH binaries that form and harden through B-B and B-S interactions.
Just over $40\%$ of the 
mergers occur within the first Gyr (about 1700 mergers), and the rate per Gyr decreases dramatically 
over time, with only about 100 mergers occurring over the last Gyr. If we assume that our models describe the 
MW GCs reasonably well\footnote{But note that our set of models does not cover realistically the 
parameter space of all MW GCs, as we show in the following section.}, we can make a very 
crude estimate of the present-day merger rate by extrapolating to the current total population of 
about 150 GCs. Multiplying our merger rate (100 Gyr$^{-1}$ during the last Gyr) for $\sim 50$ models 
by a factor of 3 (to get 150 clusters) gives us a total merger rate of $\sim 0.3$ per MWEG per Myr.

This crude estimate agrees with the merger rate that we calculated previously \citep{Morscher2013}, 
although that rate was averaged over 12~Gyr, which is clearly not reasonable considering how the merger 
 rate decreases over time. Here we use only the mergers that occurred in the last Gyr, making it 
 more appropriate for representing the current merger rate from $\sim10$ Gyr old GC systems. 
 This is comparable to the ``realistic" merger rate from primordial binaries in  
 galactic fields reported in \cite{Abadie2010}, however other more recent studies have predicted both higher
 \citep{Dominik2012} and lower \citep{Mennekens2014} field merger rates.
  Our estimate is still far too crude to accurately predict the true 
 merger rates from populations of GCs, especially since we see such extreme differences in merger 
 rate across our models. Ideally, the merger rate calculation should factor in how good of a fit our 
 models are to the MW GC population and then weigh the contribution from each model accordingly.


\subsection{Observable Properties and Comparison to Galactic GCs} \label{ResultsObservables}

In order to know whether our models are a good representation of reality we must
compare observable properties for our models to those of real Galactic
GCs. Among these key observable properties are the core radius ($r_{\rm c}$), the ratio of the 
core radius to the half-light radius ($r_{\rm c} / r_{\rm h}$), the central (3D) luminosity 
density ($\rho_0$), and the total cluster mass ($M_{cl}$). We calculate these four values
for each of our models at the final time of 12~Gyr, except for the three models that evaporated
 prior to 12~Gyr, which are not included in the following analysis.

Since the cluster mass is a straightforward quantity in our models, here we do the simplest 
thing and report our theoretical total cluster mass, which is the sum of the masses of all the 
individual stars, including dark remnants. The other three quantities are much more sensitive 
to the distribution of dark versus luminous stars, and so we must do a bit more work to obtain 
values that can reasonably be compared to the ones that observers would actually calculate. 
Since observations of GCs are generally in the $V$-band, we start by converting the bolometric 
luminosity for each star as given by BSE to $V$-band luminosities using the standard stellar 
library of \cite{Lejeune1998}. From there, the half-light radius $r_{\rm h}$ is simply the radius that 
encloses half of the light (in the $V$-band).
The core radius is a less straightforward quantity, but one that is important for its use in identifying 
the dynamical state of a GC. There are many different definitions of the core radius, and the resulting 
values can vary by a factor of a few \citep{Hurley2007, Trenti2010}. Qualitatively, the core of a cluster is the central region over 
which the density and velocity dispersion are roughly constant. More quantitatively, the core radius 
is sometimes defined as the radius at which the surface luminosity density drops to half the central 
value. To calculate $r_{\rm c}$, observers generally construct a surface brightness 
profile (SBP), and then measure where the density drops to half the central value. Alternatively, a 
King model can be fit directly to the SBP. Both of these techniques require radial binning of the 
stars, which introduces noise (since bright stars are rare) and arbitrariness (choosing a magnitude 
cutoff to remove brightest stars, choosing the bin size). In order to eliminate the need for binning 
and to smooth out the noise associated with small numbers of bright stars, we have instead opted 
to use a new and straightforward approach for calculating $r_{\rm c}$ that uses the \emph{cumulative} 
luminosity profile. To this we fit the integrated form of a King density profile, and extract the best fit 
value of $r_{\rm c}$. We find that this function provides an excellent fit to the integrated light profiles of 
our models. We describe our technique in more detail in the Appendix. 
Finally we calculate the 3D central luminosity 
density, $\rho_{\rm c}$ (in units of $L_\odot$/pc$^3$) within two different fractions of the core radius, 
$0.1\,r_{\rm c}$ and $0.25\, r_{\rm c}$. For comparison, we have also calculated $r_{\rm c}$ using the bolometric 
luminosities output by our code, and from a SBP constructed for each of our models using a technique 
similar to that in \cite{Noyola2006}. We find that the different techniques produce reasonable agreement, and 
we do not find any systematic bias in the values obtained via these three different methods.
We show the cumulative luminosity profiles and SBPs for our six representative models in the Appendix.
The observable properties for all models are given in Table~\ref{table:Observables}.

\begin{figure*} [!t]
\epsscale{0.9}

	\plotone{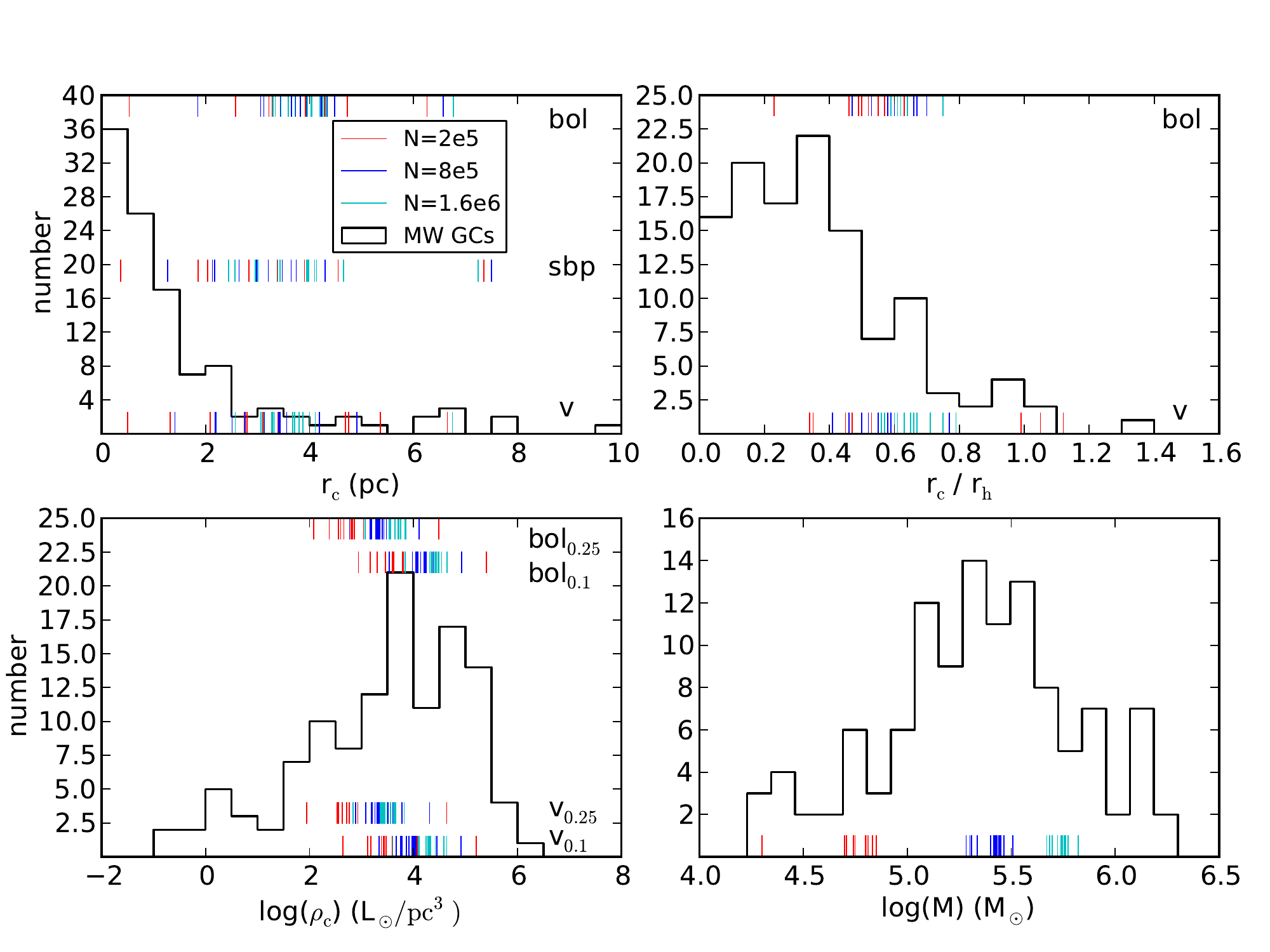}  
	\caption{\scriptsize Comparison of observable properties for MW GCs and for our models. The MW data are taken from the \citealt{Harris1996} catalog (2010 edition), excluding the masses, which are from \citealt{Gnedin1997}. The histograms show the distribution of core radii, $r_{\rm c}$, the ratio of core to half-light radius, $r_{\rm c}\, /\, r_{\rm h}$, the central luminosity density, $\rho_{\rm c}$, and total cluster mass, $M$, for the Milky Way GCs (Harris). The ticks show the calculated values of the same quantities for our models. The colors indicate the initial value of $N$. For quantities that depend on \emph{light} (all of the above, except for $M$), we have calculated the quantities with at least two different methods, which are represented by the different sets of ticks at the bottom, center, or top of the plots. For $r_{\rm c}$ and $r_{\rm h}$, the ticks on the bottom and top are calculated using the cumulative luminosity function using visual or bolometric luminosities, respectively. The ticks across the middle of the plot of $r_{\rm c}$ show the values as calculated from the surface brightness profile. For 
$\rho_{\rm c}$, the bottom set of ticks shows the luminosity density calculated in the visual band within either 0.1 $r_{\rm c}$ or 0.25 $r_{\rm c}$, and the two sets of ticks at the top of the panel represent the same quantities as derived from the bolometric luminosities. $M$ is simply the sum of all the masses in the cluster. Our clusters agree well with MW GCs in terms of $\rho_{\rm c}$ and $M$, but our measured values for $r_{\rm c}$ and $r_{\rm c} / r_{\rm h}$ fall on the high end of the distribution. The three low-$N$ models that dissolved prior to 12~Gyr are excluded from these figures.}

	\label{fig:gc_comparison}
\end{figure*}

In Figure~\ref{fig:gc_comparison} we compare $r_{\rm c}$, $r_{\rm c} / r_{\rm h}$, $\rho_{\rm c}$, and $M_{tot}$ 
for our models to the actual values observed in MW GCs. All the Galactic GC data is taken
from the \cite{Harris1996} catalog (2010 edition), except for the total 
cluster mass, which is from \cite{Gnedin1997}. The properties of the MW 
GCs are represented as histograms, and the colored tick marks indicate the final values for our 
models. Our $N=8 \times 10^5$ models produce final clusters with roughly the median GC mass 
of about $10^5\, M_\odot$, and with our three different choices for $N$, the models span most of the 
range of MW GC masses. To model the largest MW clusters, we will have to 
extend our initial $N$ to larger values. Our central densities also agree well with the bulk
of the MW clusters, with the majority of our clusters at $\rho_{\rm c} \sim 10^3-10^5\, $L$_\odot$/pc$^3$.
We miss the very high and very low density tails of the distribution, but with larger and smaller initial $N$,
and possibly other variations in our parameters, we would expect to be able to populate these regions.

\begin{deluxetable}{c@{\hskip 0.2 in}|@{\hskip 0.2 in}llll@{\hskip 0.2 in}|@{\hskip 0.2 in}llll@{\hskip 0.2 in}|@{\hskip 0.2 in}l}
\tabletypesize{\tiny}
\tablewidth{0pc}
\tablecaption{Observational quantities for all final models.}
\tablehead{
	\colhead{model} & \multicolumn{4}{c}{bolometric} & \multicolumn{4}{c}{visual} & \colhead{SBP} \\
	\colhead{} &
	\colhead{$r_{\rm c}$} & \colhead{$r_{\rm h}$} & \colhead{$r_{\rm c}/r_{\rm h}$} & \colhead{$\log_{10}(\rho_{\rm c})$} &
	\colhead{$r_{\rm c}$} & \colhead{$r_{\rm h}$} & \colhead{$r_{\rm c}/r_{\rm h}$} & \colhead{$\log_{10}(\rho_{\rm c})$} & 
	\colhead{$r_{\rm c}$} \\
	\colhead{} &
	\colhead{(pc)} & \colhead{(pc)} & \colhead{} & \colhead{($M_\odot$/pc$^3$)} &
	\colhead{(pc)} & \colhead{(pc)} & \colhead{} & \colhead{($M_\odot$/pc$^3$)} &
	\colhead{(pc)}
}
\startdata
n2w2rg2 & 2.43 & 3.18 & 0.76 & 3.59 & 1.36 & 3.13 & 0.43 & 3.77 & 2.9 \\ 
n2w2rg8 & 3.22 & 6.19 & 0.52 & 3.62 & 3.4 & 4.41 & 0.77 & 3.38 & 3.01 \\ 
n2w2rg20 & 3.34 & 6.85 & 0.49 & 3.54 & 2.09 & 5.16 & 0.41 & 3.47 & 1.85 \\ 
n2w5rg2 & 2.98 & 3.88 & 0.77 & 3.69 & 2.65 & 3.37 & 0.78 & 3.9 & 2.86 \\ 
n2w5rg8$\,\dagger$ & 3.3 & 6.54 & 0.5 & 3.6 & 4.75 & 4.24 & 1.12 & 3.59 & 3.0 \\ 
n2w5rg20 & 3.28 & 7.15 & 0.46 & 3.59 & 3.1 & 5.88 & 0.53 & 3.44 & 3.9 \\ 
n2w7rg2 & 2.92 & 3.88 & 0.75 & 3.9 & 2.36 & 3.38 & 0.7 & 4.09 & 2.13 \\ 
n2w7rg8 & 3.72 & 7.08 & 0.53 & 3.8 & 2.8 & 6.01 & 0.47 & 3.87 & 4.55 \\ 
n2w7rg20 & 4.29 & 7.82 & 0.55 & 3.3 & 2.75 & 6.14 & 0.45 & 3.18 & 2.84 \\ 
n2-A & 4.73 & 7.47 & 0.63 & 3.17 & 6.65 & 6.69 & 0.99 & 2.64 & 3.39 \\ 
n2-B$\,\dagger$ & 0.53 & 2.26 & 0.23 & 5.4 & 0.5 & 1.5 & 0.34 & 5.21 & 0.37 \\ 
n2-C & 6.26 & 10.41 & 0.6 & 2.94 & 5.36 & 9.19 & 0.58 & 3.12 & 7.35 \\ 
n2-D & 3.92 & 6.86 & 0.57 & 3.46 & 4.69 & 4.48 & 1.05 & 3.42 & 3.38 \\ 
n2-E & 2.58 & 5.43 & 0.47 & 3.79 & 1.32 & 3.77 & 0.35 & 4.07 & 2.04 \\ 
\\ \hline \\
n8w2rg2 & 3.06 & 4.96 & 0.62 & 4.38 & 2.51 & 3.99 & 0.63 & 4.45 & 2.18 \\ 
n8w2rg8 & 3.82 & 6.36 & 0.6 & 4.14 & 3.33 & 5.73 & 0.58 & 4.0 & 2.99 \\ 
n8w2rg20 & 3.95 & 6.86 & 0.58 & 4.07 & 3.13 & 6.03 & 0.52 & 3.75 & 3.65 \\ 
n8w5rg2 & 3.44 & 5.23 & 0.66 & 4.21 & 2.2 & 4.21 & 0.52 & 4.02 & 3.21 \\ 
n8w5rg8$\,\dagger$ & 3.73 & 6.25 & 0.6 & 4.2 & 2.18 & 5.35 & 0.41 & 4.29 & 2.97 \\ 
n8w5rg20 & 4.04 & 6.91 & 0.59 & 4.08 & 3.56 & 6.04 & 0.59 & 3.91 & 4.3 \\ 
n8w7rg2 & 3.65 & 5.47 & 0.67 & 4.23 & 3.41 & 4.44 & 0.77 & 4.05 & 2.13 \\ 
n8w7rg8 & 4.04 & 6.9 & 0.59 & 4.05 & 3.27 & 6.59 & 0.5 & 3.86 & 3.0 \\ 
n8w7rg20 & 4.48 & 7.45 & 0.6 & 3.98 & 4.91 & 6.39 & 0.77 & 3.67 & 3.74 \\ 
n8-A & 4.34 & 7.17 & 0.6 & 3.98 & 4.19 & 6.24 & 0.67 & 3.78 & 3.47 \\ 
n8-B & 1.85 & 3.94 & 0.47 & 4.93 & 1.41 & 3.05 & 0.46 & 4.91 & 1.27 \\ 
n8-C & 6.57 & 9.45 & 0.7 & 3.53 & 4.91 & 8.03 & 0.61 & 3.34 & 7.5 \\ 
n8-D & 4.23 & 6.65 & 0.64 & 4.05 & 3.43 & 5.66 & 0.61 & 3.97 & 4.66 \\ 
n8-E$\,\dagger$ & 3.12 & 5.88 & 0.53 & 4.25 & 2.76 & 4.99 & 0.55 & 4.29 & 2.65 \\ 
\\ \hline \\
n16w2rg2 & 3.3 & 5.12 & 0.64 & 4.65 & 3.27 & 4.34 & 0.75 & 4.58 & 2.57 \\ 
n16w2rg8 & 3.72 & 6.02 & 0.62 & 4.49 & 3.12 & 5.55 & 0.56 & 4.33 & 3.39 \\ 
n16w2rg20 & 3.96 & 6.47 & 0.61 & 4.41 & 3.3 & 5.76 & 0.57 & 4.33 & 4.1 \\ 
n16w5rg2 & 3.34 & 5.25 & 0.64 & 4.64 & 2.57 & 4.48 & 0.57 & 4.64 & 2.96 \\ 
n16w5rg8 & 4.03 & 6.28 & 0.64 & 4.43 & 3.71 & 5.64 & 0.66 & 4.24 & 3.99 \\ 
n16w5rg20 & 4.25 & 6.92 & 0.61 & 4.36 & 3.87 & 6.46 & 0.6 & 4.29 & 3.43 \\ 
n16w7rg2$\,\dagger$ & 3.59 & 5.63 & 0.64 & 4.54 & 3.07 & 4.85 & 0.63 & 4.42 & 2.44 \\ 
n16w7rg8 & 4.31 & 6.74 & 0.64 & 4.34 & 4.11 & 6.14 & 0.67 & 4.11 & 4.65 \\ 
n16w7rg20$\,\dagger$ & 4.32 & 7.07 & 0.61 & 4.31 & 3.68 & 6.52 & 0.56 & 4.09 & 4.14 \\ 
n16-A & 4.2 & 6.76 & 0.62 & 4.36 & 3.8 & 6.23 & 0.61 & 4.28 & 3.97 \\ 
n16-B & 2.07 & 4.08 & 0.51 & 5.11 & 1.59 & 3.31 & 0.48 & 5.01 & 1.65 \\
n16-C & 6.77 & 8.97 & 0.75 & 3.84 & 6.75 & 8.5 & 0.79 & 3.6 & 7.24 \\ 
n16-D & 4.05 & 6.28 & 0.64 & 4.44 & 4.0 & 5.59 & 0.71 & 4.24 & 3.95 \\ 
n16-E & 3.45 & 5.82 & 0.59 & 4.47 & 3.33 & 5.12 & 0.65 & 4.32 & 3.0 \\ 
\enddata
\label{table:Observables}
\vspace{-0.5cm}
\tablecomments{Calculations are described in Section~\ref{ResultsObservables}. 
Columns 2--5 show the core radius ($r_{\rm c}$), half-light radius ($r_{\rm h}$), $r_{\rm c} / r_{\rm h}$, and the 3D luminosity
density ($\log_{10}(\rho_{\rm c})$) calculated using the bolometric luminosities of stars as determined by BSE, 
while columns~6--9 show the same four quantities calculated using V-band luminosities (as described in the text). 
The last column shows the core radius as calculated from the SBP, also using V-band magnitudes. 
All radii are in units of parsec. The central luminosity density, $\rho_{\rm c}$, is given in units of $L_{\odot, \rm x}/\rm pc^3$, where x is either the Sun's bolometric or V-band luminosity, in the two respective calculations.}
\end{deluxetable}

We have the most trouble matching the core radius distribution of the MW GCs.
We have a deficit of models with small cores ($r_{\rm c}\, \textless\, 1$), which is where the bulk
of MW GCs fall. Only one of our models, \texttt{n2-B}, has $r_{\rm c}$ less than a parsec ($r_{\rm c}=0.5$ pc). 
This happens to be the one model that manages to get rid of nearly all its BHs, which is the 
low-$N$ cluster that starts out very compact ($R_{\rm v}=1$ pc).
The core radii for our models do span almost the full range of values occupied 
by the MW GCs, although we would still like to see more models represented
in the $r_{\rm c}\, \textless\, 1$ pc region. Instead, most of our models have core radii between about 2--5 pc.
The relatively large core radii measured also cause our $r_{\rm c} / r_{\rm h}$ values to fall on the high end 
of the distribution, although our models span a significant fraction of the range occupied by MW
clusters, except for $r_{\rm c} / r_{\rm h}\, \textless\, 0.3$.

\begin{figure}[!h]
\epsscale{0.5}	
	\plotone{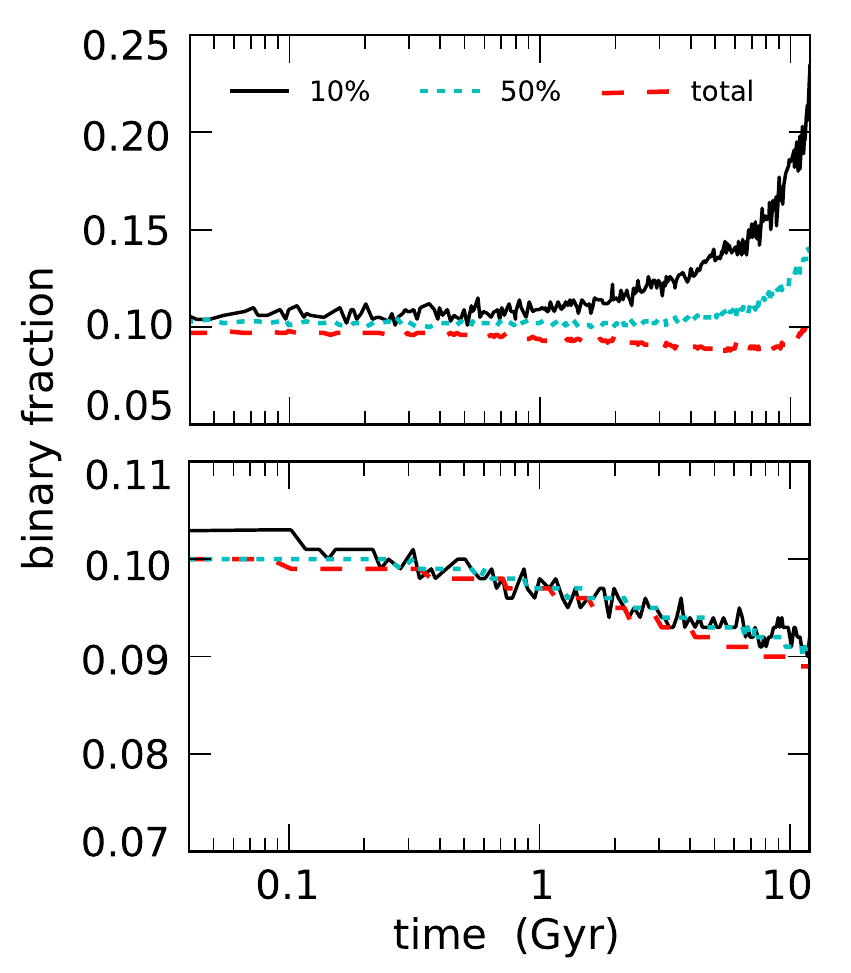}

	\caption{\scriptsize Binary fraction within the 10\% and 50\% Lagrange radii (solid black, dotted cyan, respectively)
	and the overall binary fraction (dashed red) as a function of time, for two different models, both starting with an initial
	binary fraction of 10\% (note that what we call the initial binary fraction corresponds to \emph{hard} binaries only;
	for example, for an initial hard binary fraction of 10\%, the true binary fraction would be more like 20\%).
	The top panel shows binary fractions for model
	\texttt{n2-B} and the lower panel shows model \texttt{n16w7rg20} (note the different scales on the y-axes).
	In model \texttt{n2-B} 
	the binary fraction within the 10\% Lagrange radius increases continuously with time, 
	since with the bulk of its BHs lost within a few Gyr, primordial binaries (which are much less massive than 
	typical BHs) can finally segregate into the central region of the cluster. The binary fraction within the half-mass radius
	remains fairly constant, while the overall binary fraction actually decreases slightly, until the last few Gyr, 
	when the binary fraction starts to increase everywhere. 
	Model \texttt{n16w7rg20} shows very different  behavior, with its binary fraction decreasing everywhere over
	the entire simulation. Here we do not see an increase in the inner
	binary fraction, which may have to do with heating by the significant population of BHs retained all the way to 12 Gyr, 
	which can quench the mass segregation of binaries.} 
	\emph{
	}
	\label{fig:fb}
\end{figure}

Although it is not apparent from Figure~\ref{fig:gc_comparison}, we can see in Table~\ref{table:Observables}
that there are few correlations between cluster initial conditions and final
core radii. Again we see the impact of the initial virial radius, in that the final core radius scales 
with the initial virial radius, across all models. We also notice
a slight trend of clusters with smaller $R_{\rm G}$ and higher $Z$ having slightly smaller cores. This
would be expected in models with \emph{either} a smaller $R_{\rm G}$, which are more tidally truncated, and hence
kept more compact, \emph{or} with higher $Z$, since they produce lower-mass BHs, which have less of an 
impact on the cluster, and therefore allow the cores to contract more than when more massive BHs are present. We do not find any
significant correlation between core radius and binary fraction. 
The final overall binary fractions (see Table~\ref{table:final_properties}) are, in most cases, similar to the initial values, 
while in a few of our models the \emph{core} binary fraction, $f_{\rm b,core}$, increases over time. This result is in agreement with 
numerical calculations by \cite{Fregeau2009}, and is attributed to an imbalance of mass segregation of binaries into the core,
and destruction of binaries through strong dynamical encounters (mass segregation wins out). 
In Figure~\ref{fig:fb}, we show the time evolution of the overall binary fraction and the binary fraction within the 10\% and 50\% 
Lagrange radii for two models that display contrasting behavior.
In model \texttt{n2-B} (top panel), the inner binary fraction increases steadily over time starting at around $1\,$Gyr, at
which point the cluster has already ejected about 42\% of its BHs; 85\% of the BHs are ejected by 6 Gyr.
Once the bulk of the BHs, especially the most massive ones, are lost, normal stellar binaries, which are much less massive
than typical BHs, finally begin to segregate inward. By the end of the simulation, the binary fraction within
the 10\% mass bin has more than doubled. Moreover, the 50\% and the overall binary fractions begin to increase 
in the last few Gyr. This is likely because of significant tidal stripping on this low-mass cluster
($M_{\rm cl} \approx 5.6 \times 10^4 M_\odot$ at 12 Gyr), which will preferentially remove single stars, now that the
binaries have started to segregate inward.

We do not, however, see the trend of increasing core binary fraction in all models. 
In the lower panel of Figure~\ref{fig:fb}, we show the evolution
of the binary fractions for model \texttt{n16w7rg20}, all of which actually decrease with time, even within the central
10\% Lagrange radius. In fact, in many of our larger-$N$ models, or in clusters that have longer relaxation times 
for other reasons (e.g., larger virial radius), the trend of increasing central binary fraction is less significant or not 
present at all (on the timescale of the simulations). 
We expect the timescale for segregation of binaries to scale with relaxation time, but in our models it may have 
more to do with the presence of large numbers of BHs. As discussed by \cite{Mackey2008}, the heating caused 
by a retained population of BHs can quench mass segregation of other objects (e.g., binaries) that would have 
otherwise experienced significant mass segregation within a few Gyr. This is similar to the case of an intermediate-mass BH 
quenching mass segregation by scattering stars out of the core  \citep[e.g.,][]{Baumgardt2004, Gill2008}.
As long as the BHs are dominating the central dynamics, as is the case for most of our models,
they seem to play a role in preventing segregation of binaries into the core.

Excluding the clusters that dissolved within a few Gyr, we find an overall anticorrelation between final core binary 
fraction and cluster mass (see Table~\ref{table:final_properties}), a trend that has been observed in MW GCs 
\citep{Milone2012}, as well as in the simulations of \cite{Fregeau2009} and \cite{Sollima2008}. \cite{Sollima2008} 
suggest that this could be related to the fact that cluster mass and binary destruction efficiency both have
the same dependence on cluster density and velocity dispersion. 
In our models, the trend may be due to a combination of multiple effects, including both heating (scattering) by stellar BHs 
and destruction of binaries in the core.
\cite{Milone2012} measure core binary fractions
$f_{\rm b} \lesssim 10$\% for most MW clusters observed in the study, and we find that starting with initial
binary fractions of 10\%, our final core binary fractions too remain around 10\% (typically between 9-12\%,
excluding dissolved clusters). Starting with a binary fraction of 50\% yields final binary fractions that
are much larger than those observed in GCs, but these models served as limiting cases to allow us 
to study the effect of binary fraction on the evolution of clusters with BHs.


\section{Discussion and Conclusions} \label{DiscussionConclusions}


\subsection{Summary of Results} \label{Summary}

Our goal here was to study the evolution of massive star clusters that initially retain most of their BHs
in order to see whether it is possible for many BHs to remain after $\sim10\,$Gyr
and still have cluster properties consistent with those of MW GCs. Most of our clusters
do indeed retain many BHs at the end of the simulations (up to $\sim10^3$, for initial $N$ from 
$2\times 10^5-1.6\times10^6$), but the agreement with observable properties of MW GCs is not perfect.
Qualitatively, all of our models evolve quite similarly, at least during the first few Gyr.
The BHs quickly become very centrally concentrated, but, as a
whole remain spread out over about a parsec in radius, similar to the innermost $10\%$ of the 
non-BH mass. At the very center, the most massive BHs drive repeated core oscillations 
where a few tens of BHs collapse into a cusp, but then promptly re-expand via their 
own dynamics after forming three-body binaries. 
Single and binary BHs are ejected over time, with the most massive BHs being ejected first,
followed by the less massive ones. 
As this happens, the remaining population of lower-mass BHs
becomes less efficient at driving deep core collapses and the outer envelope of 
the oscillating central $1\%$ BH mass slowly expands.
This results in a lower central density and a gradual slowing of both the interaction 
rate and the BH ejection rate.
While most models still have a significant population of $5-15\, M_\odot$ BHs at the end,
we find that this depends sensitively on the initial conditions,
in particular quantities that have the potential to significantly modify the cluster relaxation 
time, such as $N$ and $R_{\rm v}$.
Clusters with shorter relaxation times (lower $N$, smaller $R_{\rm v}$) process their BHs more 
quickly, and therefore end up retaining smaller fractions of their initial BH populations by 12 Gyr. The mass 
of the most massive bound BH depends on the extent to which the BHs have been depleted. 
Model \texttt{n2-B} has the smallest $N$ ($2 \times 10^5$) and $R_{\rm v}$ (1 pc), and therefore 
the shortest relaxation time of all, and it retains just 9 BHs at 12~Gyr (about 2\% of the initially 
retained BHs). Similarly, models \texttt{n8-B} and \texttt{n16-B} (larger $N$, but $R_{\rm v}=1$ pc) retain the fewest BHs 
among other models with the same $N$. 

Our models have final binary fractions that agree very well with observation. 
The total masses and luminosity densities for our models also fit well within the 
parameter space of observed MW clusters. Our final core radii, however, are for the most part too 
large to represent the bulk of MW clusters, although they do fall along the extended tail of 
the distribution. Like the dynamics of the BHs, the final core radii too are affected
significantly by the initial virial radius. All models with $R_{\rm v}=2$ pc have 
final core radii between $3-5$ pc, and those with $R_{\rm v}=4$ pc 
have even larger cores ($6-7$ pc), regardless of $N$. 
The only models to eventually contract down to core sizes smaller than 2~pc are the 
three that \emph{start} much more compactly with $R_{\rm v}=1$ pc.
These models all reach 
a point at which the BHs are providing so little energy that the cluster as a whole stops expanding,
or in the case of model 
\texttt{n2-B}, actually starts contracting (see Figure~\ref{fig:lagrad}), as the remaining
low-mass BHs start to lose their dominance at the cluster center.
In the last few Gyr of evolution in model \texttt{n2-B} the BHs become more and more 
integrated with the rest of the cluster, and finally the cluster core (i.e., the \emph{observational} 
core, composed of luminous stars) starts to contract, resulting in a final core size of just 0.5 pc. 
This model has a mass of only $2 \times 10^4\, M_\odot$ at 12~Gyr, placing it at the very bottom 
of the MW GC mass distribution. In fact, the three most compact clusters lose mass at a faster rate 
overall than the comparable model with larger $R_{\rm v}$, and also end with fewer BHs and smaller cores, 
indicating that BH evaporation seems to be tied very closely to \emph{cluster} evaporation.
However, models \texttt{n8-B} and \texttt{n16-B} each have about the same final mass as the model 
\texttt{n8w5rg2} and \texttt{n16w5rg2}, respectively (which have $R_{\rm v}=2$ pc, but smaller tidal radii)
yet being more compact, they still eject their BHs more efficiently and therefore achieve smaller core sizes in the end.
 
Each of our models forms and ejects many BH binaries over the course of their evolution,
but the majority of the BHs, both retained and ejected, are single BHs.
Most of the ejected binaries are BH--BH, but some BH--non-BH binaries are ejected as well. 
The number of ejected BH binaries, their properties at ejection, and therefore the number of subsequent BH--BH mergers
(inside and outside of the clusters) depend primarily on $N$ and $R_{\rm v}$. 
We produce many BH--BH mergers (more than 4000 in total), with at least one merger produced in each cluster. 
Roughly 60\% of the ejected BH--BH binaries actually merge within a Hubble time (we do not calculate 
this fraction for retained BH--BH binaries because their properties are still being modified by dynamics).
Since our models do not yet show great agreement with all relevant properties of Galactic GCs, we cannot yet
make any reliable quantitative predictions about the numbers of interesting BH binary systems in our Galaxy, 
or other similar galaxies.


\subsection{Uncertainties and Comparison to Other Studies} \label{Comparison}

\cite{Breen2013} were the first to suggest that the dynamics of a population of BHs is
actually regulated by the \emph{cluster}, and that for this reason BHs can be retained for much 
longer than previously thought. In their simplified two-component cluster models they found
that the BHs behave such that they meet the energy needs of the cluster, similar to the way that 
primordial binaries balance energy flow during the binary-burning phase \citep{Fregeau2007, Gao1991}. 
Earlier studies \citep{Mackey2008, Merritt2004} have demonstrated that the interactions 
(and subsequently the ejections) of a segregated population of BHs can inject enough heat 
to cause significant core expansion in clusters.  
\cite{Breen2013} find that some point, however, there are too few BHs to balance the energy 
lost via relaxation, and only then can the cluster finally approach the phase that the authors 
call \emph{second core collapse}, to distinguish it from the initial BH-driven collapse 
(\emph{second} core collapse therefore refers to what is usually just called \emph{core collapse}).
If the \emph{cluster} drives the rate of energy flow, we should not expect the BHs to evaporate
within a few cluster relaxation times.
Our results agree with this basic picture, and we see this very behavior play out in our 
model \texttt{n2-B}, which is actually approaching the second core collapse phase by the end of the simulation. 
All of our models display many deep collapses of a small number of BHs, but the formation of 
three-body binaries and their subsequent interactions ultimately power the re-expansion of 
the cusp. We find that the BHs actually spend \emph{most} of their time in the \emph{uncollapsed} state,
which also helps to explain how it is possible for BHs to remain in clusters for so long. The BHs \emph{try} to 
decouple via the Spitzer instability, but their own dynamics ensures that they always re-couple to
the cluster very quickly.

\begin{figure} [!t]
\epsscale{0.7}

	\plotone{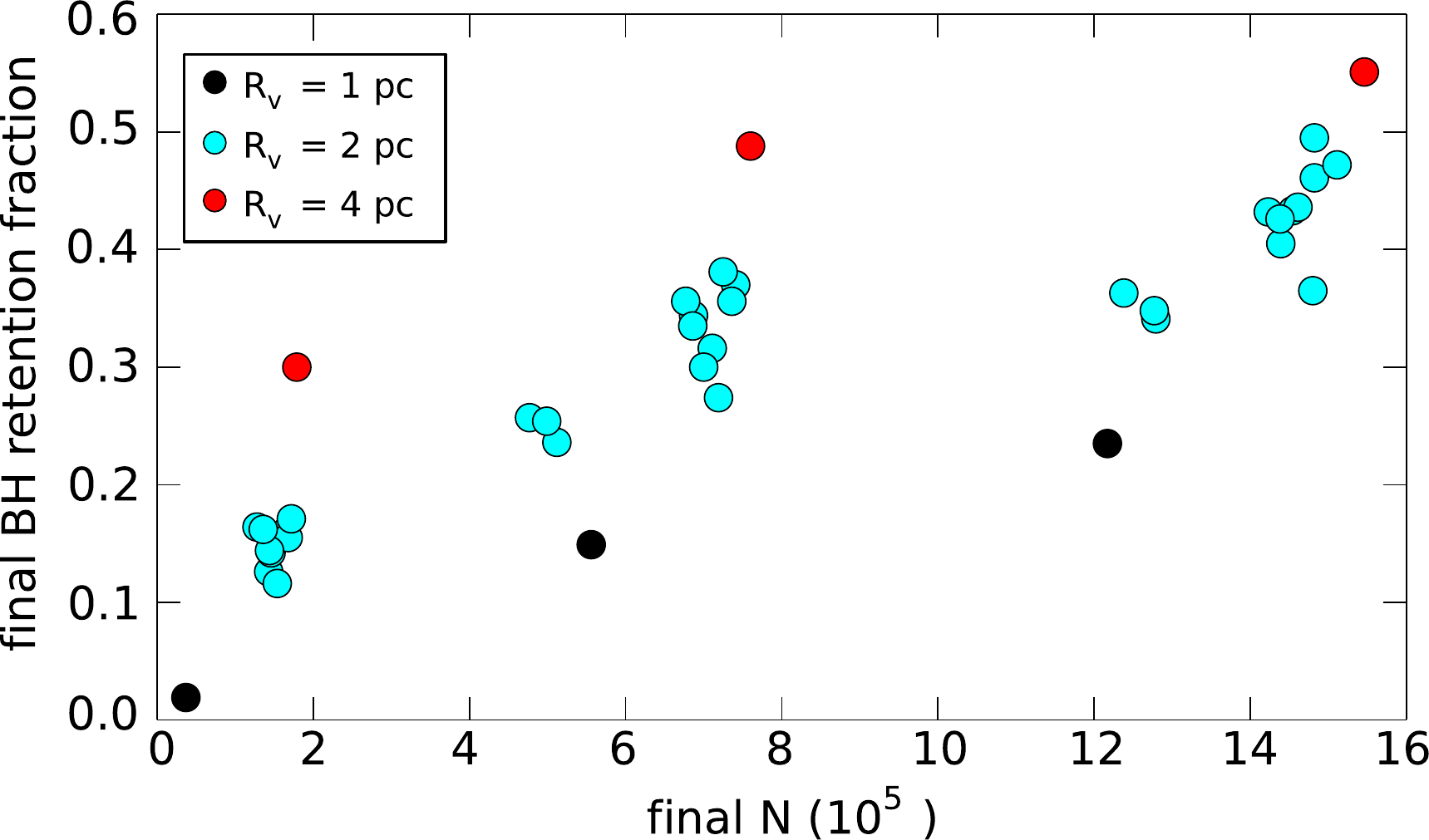}
	\caption{\scriptsize Relationship between final BH retention fraction and final $N$ for all models that survive to 12~Gyr.
	The different colors indicate the initial virial radius.
	The trends of increasing BH retention fraction with $N$ and with $R_{\rm v}$ are expected
	since the relaxation time depends on $N$.
	Clusters with either smaller $N$ or smaller $R_{\rm v}$ have shorter evolutionary timescales, 
	and are therefore in a later dynamical state at 12~Gyr, so they have ejected larger 
	fractions of their BHs, as well as lost more mass overall. 
	}
	\label{fig:bhfrac_vs_n}
\end{figure}

\cite{Heggie2014} also discuss the dependence of BH retention on relaxation time by
comparing models of four clusters with very different initial conditions (modeled after M4, NGC~6397, 47~Tuc and M22).
The M4 and NGC~6397 models both have short relaxation times, and although they retain nearly all the BHs \emph{initially}, 
they eject almost all of them within 12~Gyr. In contrast, the 47~Tuc and M22 models start with only 10\% of the formed BHs
initially, yet given their longer relaxation timescales most of these BHs still remain at 12~Gyr.
The difference, they explain, is that the models are in different dynamical states: the former two models have reached
the second core collapse phase, while the latter two are far from it.
They predict that clusters with long relaxation times are more likely to still contain many BHs at present.
We have shown that this is true for our models as well, but to make the point more clearly we show in Figure~\ref{fig:bhfrac_vs_n} the 
relationship between the final fraction of BHs retained and the final number of bound stars for
our models. This shows that the trend predicted by \cite{Heggie2014} holds roughly for a wide range 
of initial conditions, although the variety of initial conditions also leads to the large amount of scatter. 

While these studies (as well as our own) agree that BH dynamics is regulated for the most part by the 
cluster, there are still uncertainties about initial BH populations in GCs that may impact the precise 
evolutionary timescale for BH evaporation, and therefore predictions for present-day clusters.
The BH mass spectrum (derived from the remnant-to-progenitor mass relationship and the upper end of the stellar IMF) 
is somewhat uncertain. Since the most massive BHs are ejected first, and the impact of BHs on 
the cluster lessens with time as BHs are ejected, the BH mass spectrum could significantly affect the long-term evolution 
of BHs and clusters.
Another key uncertainty is the magnitude of birth kicks for BHs.
Studies attempting to constrain BH kick strengths using observations of BH XRBs have led to mixed conclusions. 
\cite{Repetto2012} suggested that large kicks (similar to those of NSs) were necessary to explain 
the spatial distribution of BH XRBs in the Galaxy, while other studies have found that lower natal 
kick velocities better explain the properties of at least some specific systems \citep[e.g.,][]{Willems2005,Wong2012}.
There are also competing theories for the origin of these kicks (e.g., neutrino-driven versus supernova-driven)
which lead to different predictions for the magnitudes that we should expect for BHs (see summary in \citealt{Janka2013}).
\cite{Janka2013} presented a new kick model that might explain how BHs could acquire kicks
similar to those of NSs. The model suggests that asymmetric supernova ejecta could lead to an acceleration of the remnant BH 
gravitationally in the same direction as the initial kick, in which case the kick momentum 
\emph{grows} with BH mass. 
Recent models of M22 presented in \cite{Sippel2013} and \cite{Heggie2014} found good agreement with observable properties of M22 by starting with very small initial retention fractions (10\%, or about 50 BHs), under the assumption that BHs receive the same kicks as NSs.
For comparison, we have repeated 3 simulations (\texttt{n2w5rg8}, \texttt{n8w5rg8}, and \texttt{n16w5rg8}), except 
we allowed BHs to receive kicks identical to those of NSs (kick chosen independent of mass or fallback). As expected, we find 
that very few BHs are retained initially (0, 6, and 31, for the three models, respectively), and they have little
effect on their host clusters. It seems that such a small number of BHs cannot power the deep collapses that we 
have seen in the simulations described in this work. This also helps to explain how these models that retained just 6 and 31
 BHs initially, still managed to keep 5 and 19 of them (respectively) all the way to 12 Gyr, since BH ejections tend to 
 occur predominantly during the deep collapse phases, which are absent from the models with very few BHs.
However, since the goal of this work is to better understand the evolution of clusters that retain most of
their BHs initially, and to determine to what extent these clusters resemble our Galactic GCs, 
we have chosen to focus on the effects of varying \emph{only} the initial conditions of the cluster models as a whole, 
and have left the initial BH populations fixed, except for the differences that arise naturally from 
different choices for cluster parameters. The effect of BH kicks and the BH mass spectrum will be the topic of a future study.

With the growing evidence for BH XRBs in old GCs, it would also be interesting to use our models to 
predict the numbers and properties of BH XRBs in GCs.
However, we cannot trust our simple treatment of binary stellar evolution to predict 
the behavior of these binaries. This would require more focused binary evolution and mass transfer modeling of specific systems that form in our cluster models. A very crude analysis of the entire population of retained 
BH binaries with non-compact companions at 12 Gyr indicates that at least some of these systems 
(28, or about 13\%) could potentially be interesting X-ray sources at present.
Since we use the standard ``sticky sphere" approximation for physical collisions, our code is also not 
capable of predicting detailed outcomes of collisions between BHs and non-compact stars. Our standard treatment leads
effectively to the entire mass of the colliding objects to be entirely and immediately accreted onto the BH, ignoring completely
any feedback effects (which could lead to significant mass loss) or the finite timescale of the accretion flow. 
Our code can, however, predict the \emph{rates} of these collisions and as a quick test we have checked 
how many collisions occur between BHs and non-degenerate stars in one of our large-$N$ models (\texttt{n16w5rg8}).
In total there were 45 direct collisions involving a BH and a non-degenerate star, which occurred via different kinds of 
interactions: 23 occurred as direct
S-S collisions, 18 during strong B-S interactions, and 4 during
B-B interactions. Most of these collisions were with main-sequence stars (40),
but there were also a few collisions with giants (5), which, if treated in more detail would have likely led 
to the formation of a compact BH--WD binary remnant \citep{Ivanova2010}. About half of the collisions happen within
the first Gyr, and the rate declines after that. We also see 21 evolutionary mergers between a
BH and a non-degenerate star, which occurred during binary stellar evolution rather than during
dynamical encounters. All of these mergers were with MS stars, and they all occurred 
within the first 13~Myr. We have not yet studied the details of these collisions and mergers, but
it would be interesting to look at their properties, such as impact parameter and stellar masses,
and then predict the possible observable outcomes of such events, as they could produce 
transient sources that would be detectable by surveys such as LSST \citep{LSST2009}.

Perhaps most importantly, on the computational side, since we find that cusps involving a few 
tens of BHs form repeatedly in our models, we must ask whether an orbit-averaged MC 
approach can model this dynamical behavior accurately. 
In particular, for a small-$N$ decoupled subsystem, the relaxation and dynamical 
timescales can become comparable, in which case the Fokker-Planck approximation, 
a key assumption in the MC technique, breaks down.  
Furthermore, such a small number of particles makes
the estimation of local averages highly susceptible to Poisson fluctuations,
which directly influences the accuracy of all dynamical calculations.
The direct $N$-body technique does not suffer from these issues, and so it can handle 
the dynamics of a small-$N$ system quite naturally. 
This is most likely responsible for the difference in core radius between the two
methods that was noted in Figure \ref{fig:2species64k}.
In a BH-driven
collapse, the small number of massive particles interact on a much shorter
timescale than our relaxation timestep can resolve.  In order to address this, we
are currently developing a new technique that will allow us to model the
dynamics of these deep collapses more accurately.  This hybrid $N$-body/Monte Carlo
technique integrates the dynamics of the BHs (or other massive
particles) directly with an $N$-body integrator, while
the majority of lower-mass stars in the halo and core interact via the two-body
relaxation of the MC approach.  Preliminary results indicate that this
technique achieves similar speed to a pure MC simulation while
producing
core radii that agree with results observed in full $N$-body
simulations \citep{Rodriguez:2014ip}.


\subsection{Conclusions} \label{Conclusions}

Starting with reasonable initial conditions describing young star clusters we have presented many simulations of GCs
containing populations of hundreds to thousands of BHs. Without any fine tuning of parameters, we find that our models have present-day 
observable properties that are consistent with the MW GCs, although our core radii are slightly large. 

Our main conclusion is that \emph{if} most BHs are retained initially, it seems that the only
 way to still eject most or all BHs by $\sim12\,$Gyr is to start with very compact clusters. If clusters can eject
 enough BHs, then the core can finally begin to contract, producing final core radii that may be
in better agreement with those observed in MW clusters.
Most of our models, on the other hand, retain significant numbers of BHs all the way to 12~Gyr (typically
$\approx 50-100$ for our lowest-$N$ models, and $\approx 1000-2000$ for our
largest-$N$ models), and have rather large cores (typically about 2--5 pc).
We confirm that the BH evaporation timescale is set by the cluster evolutionary timescale as suggested 
by \cite{Breen2013} and \cite{Heggie2014}. We find that the BHs drive deep core oscillations during 
which a small number of BHs can form a steep cusp, but these always re-expand and re-mix with the 
other stars very quickly, and the result is that most of the time the BHs are in their uncollapsed state, 
well mixed with other stars. We suggest that this may explain why the BHs mostly avoid the Spitzer instability, 
and hence why they can be retained for much longer timescales than previously thought.
 
It will be important to test the effect of uncertain stellar parameters, especially those pertaining to the 
BH populations, such as BH birth kick magnitudes and the BH mass -- progenitor mass relationship, 
which we have not explored in this study.  These parameters will undoubtably affect the subsequent 
dynamics of the BHs and the clusters as a whole, and may therefore also change the predictions for 
BH retention. If it turns out that BHs do indeed get kicks of the same magnitude as NSs and so at most 
only $\sim10$\% are retained initially, then the very compact initial conditions might not be necessary 
to produce small cores by $\sim12$~Gyr, since there would be far fewer BHs to eject before the core 
could start to contract. 

 In order to derive the proper BH--BH 
 merger rate for MW-equivalent galaxies (and the corresponding predicted LIGO detection rate)
 we will first need to run additional simulations in order to fill in the gaps where we currently have poor 
 coverage in the parameter space of observed MW GCs. 
 We will then be able to do a detailed statistical calculation that weighs the contribution from each of our models according to how well their 
 properties match the MW population. This calculation will be the topic of a forthcoming paper (Rodriguez et al., in preparation).


\clearpage
\appendix
\section{Calculation of Observational Core Radius}
To make a SBP requires the stars to be binned radially. The bins should be small enough in radius that the core is resolved. The tradeoff is that small bin sizes increase the random noise, since a single bright star can dominate the light
for an individual bin, introducing large bin-to-bin variations. To get around this, observers generally remove the brightest stars
before calculating the SBP \citep{Noyola2006} based on a somewhat arbitrary choice of a magnitude cutoff. A detailed discussion and comparison of  various techniques can be found in \cite{Noyola2006}.

To avoid the complications of binning, we have chosen to use a new technique for calculating the core radius.
Our technique involves fitting a king model to the \emph{cumulative} luminosity function, which is much smoother 
than the luminosity density (or surface brightness profile) because it does not require us to bin the stars.  
We start with the analytic approximation to the King model density profile (Equation~13 from \citep{King1962}),
\begin{equation}
\Sigma(r) = \frac{\Sigma_o}{1+{(r/r_{\rm c})}^2},
\label{eq:SigmaKing}
\end{equation}
where $\Sigma_o$ is the central 2D surface density and $r_{\rm c}$ is the King core radius. We then integrate this 
equation over the surface area out to some distance $r$, so that it is now represents the \emph{cumulative} 
luminosity as a function of $r$,
\begin{equation}
L_{tot}(r) = \pi \Sigma_o r_{\rm c}^2 \log \left( 1 + (r/r_{\rm c})^2  \right).
\label{eq:IntegralSigmaKing}
\end{equation}
Finally, we fit this equation to the cumulative luminosity profile for each of our models to find the best values 
for $\Sigma_o$ and $r_{\rm c}$.  

\begin{figure*} [!ht]
\epsscale{0.8}

	\plotone{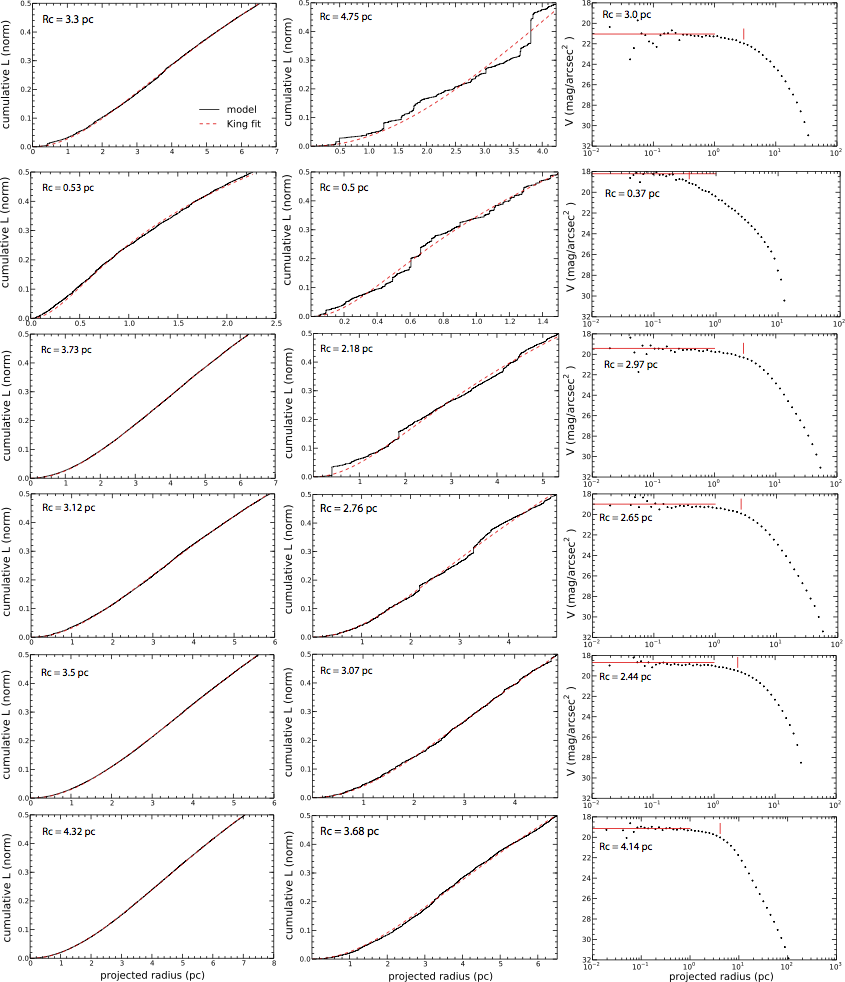}  
	\caption{\scriptsize Core radii calculations for the six models shown in Figure~\ref{fig:lagrad}.
The left panels show the cumulative luminosity profile calculated from the
bolometric luminosities (solid black curve) and the King fit to the model (red dashed curve). 
The resulting $r_{\rm c}$ obtained using our new technique (as described in Section 
\ref{ResultsObservables} and the Appendix) is given on each panel. The center panels 
show the same thing, but using V-band luminosities. 
On the right panels we show the V-band SBP for each model
with a vertical red tick mark to indicate the location of $r_{\rm c}$, the point at which
the surface luminosity density drops to half the central value. The horizontal red line 
indicates the central brightness. For simplicity we assume all 
clusters are at a distance of 8.5 kpc. 
This choice does not affect the core radius measurement, but it does affect the magnitude 
scale (y-axis), and therefore the numerical values here are somewhat arbitrary.}
 
	\label{fig:core_radii}
\end{figure*}

All core radii given in columns~2 and~6 of Table~\ref{table:Observables} are calculated using this technique,
based on either the bolometric or the V-band luminosities. We show the cumulative luminosity profiles and the 
SBPs for a sample of six of our models in Figure~\ref{fig:core_radii}.

\acknowledgements

We thank the anonymous referee for several suggestions that improved this work.
This work was supported by NSF Grant AST--1312945 and NASA Grants NNX09AO36G 
and NNX14AP92G at Northwestern University. MM acknowledges support from an 
NSF GK-12 Fellowship funded through NSF Award DGE- 0948017 to Northwestern University. 
FAR acknowledges the hospitality of the Aspen Center for Physics, supported by NSF Grant PHY-1066293. 
All computations were performed on Northwestern University's HPC cluster Quest.

\clearpage

\bibliographystyle{hapj} 

\end{document}